\documentclass[a4paper,11pt]{article}
\usepackage{jheppub} 
\usepackage{graphicx}
\usepackage{float} 
\usepackage{xcolor}
\usepackage{amsmath, amssymb}
\usepackage{subcaption}
\usepackage{comment}
\usepackage{cancel}
\usepackage[normalem]{ulem}

\newcommand{\BB}{\ensuremath{B_g}}

\title{\boldmath  Connecting boundary entropy and effective central charge at holographic interfaces}

\author[a]{Evangelos Afxonidis,}
\author[a]{Ignacio Carreño Bolla,}
\author[a]{Carlos Hoyos,}
\author[b]{and Andreas Karch}

\affiliation[a]{Department of Physics and Insituto de Ciencias y Tecnolog\'ias Especiales de Asturias (ICTEA), Universidad de Oviedo, c/ Leopoldo Calvo Sotelo 18, ES-33007, Oviedo, Spain}
\affiliation[b]{Weinberg Institute, Department of Physics, University of Texas, 2615 Speedway, Austin, TX 78712, USA}

\emailAdd{afxonidisevangelos@uniovi.es}
\emailAdd{ignaciocarbolla@gmail.com}
\emailAdd{hoyoscarlos@uniovi.es}
\emailAdd{karcha@utexas.edu}

\abstract{The entanglement entropy of intervals in $1+1$ interface CFTs is modified in two ways compared to a CFT without interface: there is a finite boundary entropy contribution, and, for an interval with an endpoint at the interface, the coefficient of the logarithmically divergent contribution -which is usually proportional to the central charge of the CFT- is modified to an effective central charge.  We show that the latter modification can be understood as a limit of the former using holographic duals of interface CFTs. Furthermore, we show that a finite contribution also appears in intervals that do not cross the interface and it is needed to ensure strong subbaditivity of the entanglement entropy.}

\begin{document}
\maketitle
\flushbottom

\section{Introduction}\label{sec::intro}
Interface conformal field theories (ICFTs) \cite{Bachas_2002} and boundary conformal field theories (BCFTs) \cite{Cardy:1986gw} have garnered significant interest in recent years and apply to both condensed matter systems and string theory. Here we will mainly focus in the lesser explored ICFT case, describing a system where a conformally invariant interface connects two, potentially distinct, conformal field theories (CFTs) with differing central charges. We will restrict our discussion to $1+1$ theories with a codimension-one interface localized in space.

ICFTs give rise to novel entropic and energetic quantities, such as the interface effective central charge and entropy function \cite{Karch_2021,Karch_2023,karch2023universalityeffectivecentralcharge,karch2024universalboundeffectivecentral,Afxonidis:2024gne} and the energy transmission coefficient $\mathcal{T}$ \cite{Quella_2007,Meineri_2020}, all three of which are inherently defined in the presence of interfaces and do not arise in boundary CFTs. In this work, while we briefly comment on the role of the transmission coefficient, our primary focus will be on the entanglement entropy (EE) associated with a spatial interval $A$, which may either contain the interface or lie entirely on one side of it. 

The EE of an interval $A$ containing the interface, and symmetric around it, can be computed using the replica method \cite{Calabrese_2009} as
\begin{equation}\label{eq::EEBCFT}
    S_A=\frac{c_L+c_R}{6}\log{\frac{l}{\epsilon}}+\log g \ ,
\end{equation}
where $l$ is the total length of the interval, $\epsilon$ is a UV regulator and  $c_L$ and $c_R$ are the central charges of the two CFTs around the interface. The entropic quantity $\log g$ appearing in \eqref{eq::EEBCFT} plays a central role in the characterization of defects and boundaries, and is commonly referred to as the boundary entropy number \cite{Calabrese_2009}. It was originally introduced from a thermodynamic perspective in \cite{PhysRevLett.67.161}, and captures the number of degrees of freedom localized at a defect, serving as a boundary analogue to the central charge $c$ in a CFT. While exact results for $\log g$ are available in many free or integrable models, its evaluation in strongly coupled systems remains a significant challenge. Holography offers a valuable framework: the boundary entropy can be accessed via the holographic entanglement entropy (HEE) prescription \cite{Ryu_2006}. We will use holography to analyze the EE of ICFTs using a general construction where the geometry admits a foliation in $AdS$ slices, and consider several specific examples: RS braneworlds \cite{Karch_2001,Karch_2001_2}, Janus \cite{Bak_2003} and super Janus geometries \cite{Chiodaroli_2010}.

We will study some of the interesting features that emerge when considering an asymmetric interval around the interface \cite{Karch_2021,Anous_2022}. In this case, the boundary entropy is generalized to an interface entropy function $\log g_{\text{eff}}$. The value of the interface entropy function depends on the distances of the endpoints of the interval to the interface, $l_L$ and $l_R$ for the leftmost and rightmost points respectively, with conformal invariance dictating the dependence to be through their ratio $l_L/l_R$. In the limit where $l_L=l_R$ the interface entropy function takes the value of the boundary entropy number $\log g_{\text{eff}}=\log g$ by definition. In this work we will extend the definition of the interface entropy function to intervals that do not contain the interface. The interface entropy in this case is also a function of the ratio $l_L/l_R$, and, as we will show, it gives rise to non-trivial contributions to the EE that are needed to preserve Strong Subadditivity. 

A different observation concerning asymmetric intervals was that when one of the endpoints of the interval is located at the interface, $l_L=0$ or $l_R=0$, the coefficient of the logarithmic divergence is modified, replacing $c_L$ or $c_R$ by a different value, dubbed as effective central charge $c_{\text{eff}}$ \cite{Sakai:2008tt,Brehm:2015lja,Karch_2001,Karch_2023,karch2023universalityeffectivecentralcharge,karch2024universalboundeffectivecentral}.  The effective central charge has emerged as a key concept in recent developments, where it has been interpreted as a measure of quantum information transmission across the interface.

Notably, $c_{\text{eff}}$ satisfies a set of intriguing bounds in relation to the transmission coefficient $\mathcal{T}=2 c_{LR}/(c_L+c_R)$, namely $0\leq c_{LR}\leq c_{\text{eff}}\leq \min (c_L,c_R)$ \cite{karch2023universalityeffectivecentralcharge,karch2024universalboundeffectivecentral,Bachas:2022etu}. We will show in this work that the effective central charge and the interface entropy are directly related, the former emerges from a limit of the latter. Thus the interface entropy provides a unified description of the EE in interfaces, for all interval layouts.

The paper is organized as follows. In Section \ref{sec::reviewJanussuperJanus}, we review the general holographic dual description of ICFTs and the specific examples mentioned before.  In Section  \ref{eq::secEEcomputations} we review the holographic prescription to compute the entanglement entropy and obtain the interface entropy for intervals with and without the interface in the interior. In section \ref{sec::validitySSA} we study the Strong Subadditivity condition in view of the monotonicity properties of the interface entropy. In section \ref{sec::connectionloggceff} we show the connection between interface entropy and the effective central charge and derive some related results. In section \ref{sec::unequalcentralcharges} we extend the analysis to asymmetric ICFTs and to BCFTs. We end in section \ref{sec::discussion} with a discussion of our results and possible future directions.

\section{Holographic duals of  
ICFTs}\label{sec::reviewJanussuperJanus}

In this section we will describe general features of holographic duals of $1+1$-dimensional ICFTs, and review some specific examples that we will employ throughout the paper to illustrate our results. The examples are the subcritical Randall-Sundrum (RS) braneworld \cite{Karch_2001,Karch_2001_2},  Janus solution \cite{Bak_2003} and the super Janus solution \cite{Chiodaroli_2010}.

Holographic duals of $1+1$-dimensional ICFTs are frequently described by three-dimensional geometries. From the standpoint of string theory, consistency requires that such lower-dimensional geometries arise as consistent truncations of ten-dimensional spacetimes, typically involving compactification on a seven-dimensional internal manifold \cite{Chiodaroli_2010,baig2024transmissioncoefficientsuperjanussolution} . In many instances, this internal space is non-trivially fibered over the non-compact part of the geometry. The Janus and super Janus solutions fall into this category. Bottom-up constructions where the ICFT is directly constructed in three dimensions can also be considered, with the RS construction being the simplest instance of this class.

In all these cases, the three-dimensional metric is usually written using an $AdS_2$ slicing of the full geometry. Taking $r$ as a domain wall coordinate
\begin{equation}\label{eq::defectmetric}
    ds^2=R^2 \left(dr^2+e^{2A(r)}\frac{dx^2-dt^2}{x^2}\right) \ ,
\end{equation}
the $(t,x)$ coordinates span an $AdS_2$ subspace with $x=0$ the asymptotic boundary. For a warp factor $e^{A(r)}=\cosh r$ the metric above is actually an $AdS_3$ geometry of radius $R$. Indeed, the following change of coordinates
\begin{equation}\label{eq:poincarecoord}
    z=\frac{x}{\cosh{r}}\ , \quad y=x\tanh{ r } \ ,
\end{equation}
maps the metric to the usual Poincar\'e patch coordinates
\begin{equation}\label{eq::Poincarepatch} 
    ds^2=\frac{R^2}{z^2} \left(dz^2-dt^2+dy^2\right) \ .
\end{equation}
The boundary is at $z=0$ with $y$ spanning the spatial direction. We can reach each half of the boundary from the bulk by taking $r\to +\infty$ ($y>0$) or $r\to-\infty$ ($y<0$). Taking $x\to 0$ one reaches the boundary at $y=0$. Thus $y=0$ is a distinguished point in the $AdS_2$ slicing and there is a codimension one defect or interface sitting at this point for generic choices of the warp factor. In general, the warp factor will reach a minimal value $e^{A_*}$ at some $r=r_*$ that will play a significant role in computations of the entanglement entropy, studied in Section \ref{eq::secEEcomputations}.

\subsection{The RS braneworld}

An illustrative and extensively analyzed example (see for example \cite{Karch_2021,Afxonidis:2024gne} for a study of the interface entropy), is the subcritical Randall-Sundrum (RS) braneworld \cite{Randall_1999_1,Randall_1999_2,Karch_2001,Karch_2001_2}. This construction consists of  Einstein gravity with a negative cosmological constant and sourced by a brane-like matter configuration.
The brane itself is modeled as an thin, relativistic hypersurface endowed with constant energy density and parametrized by its tension $T$. When the brane tension is below a critical threshold, it intersects the asymptotic $AdS$ boundary. The bulk regions are described by pure $AdS$ geometries glued at the location of the brane. Such configuration can be naturally proposed as a bottom-up holographic realization of an ICFT, albeit without a concrete embedding in string theory. This type of construction also appears naturally in the description of impurities in higher dimensions after a $s$-wave reduction \cite{Erdmenger:2013dpa,Erdmenger:2015spo,Erdmenger:2015xpq}.

In our setup, the RS braneworld can be understood as a specific realization of the warp factor
\begin{equation}
    e^{A}=\cosh{\left(|r|-r_*\right)} \ ,
\end{equation}
where the warp factor is piecewise defined for positive and negative $r$. There is a discontinuity of the derivative at $r=0$, which indicates the presence of the brane there . A crucial distinction from the pure $AdS$ case is that the minimal value of the warp factor $e^{A_*}=1$ occurs at $r=\pm r_*$, rather than at $r=0$. 

The brane tension $T$, the turnaround point $r_*$, 
and the boundary entropy $\log g$ are related as \cite{Takayanagi:2011zk}
\begin{equation}\label{eq::boundaryentropyRS}
    \log g=\frac{2 R r_*}{4G}=\frac{cr_*}{3}=\frac{c}{3}\tanh^{-1}{\left(\frac{T}{2}\right)} \ ,
\end{equation}
where $c$ is the central charge of the CFT and we have used the Brown-Henneaux relation \cite{Brown:1986nw}
\begin{equation}\label{eq:centralc}
    c=\frac{3R}{2G}\ .
\end{equation}

As we will discuss in more detail in Section \ref{sec::unequalcentralcharges}, this construction can be extended to interfaces separating CFTs of unequal central charges, and BCFTs in the limit when the central charge of the CFT on one side vanish.

\subsection{Janus}\label{sec::reviewJanus}

We now consider the Janus solution, originally introduced as an aymptotically locally $AdS_5\times S^5$  supergravity dual of a 3+1 dimensional ICFT \cite{Bak_2003}. A 1+1 dimensional realization has a supergravity dual which is asymptotically locally $AdS_3\times S^3\times T^4$ \cite{Bak_2007}. The ten-dimensional metric is given by
\begin{equation}
    ds^2_{10}=e^{\phi/2}\left(ds^2_3+d\Omega^2_3 \right)+e^{-\phi/2}ds^2_{T^4} \ .
\end{equation}
Note that the standard Janus solution comes with a three-dimensional metric that appears as a direct product factor in the ten-dimensional metric. This makes the geometry of the standard Janus solution significantly simpler than its supersymmetric counterpart, discussed in Section \ref{sec::reviewsuperJanus}. 

The reduced three-dimensional theory is described by the Einstein-Hilbert action plus the dilaton $\phi$  \cite{Chiodaroli_2010}. The metric for the Janus solution takes the form in \eqref{eq::defectmetric} with a warp factor
\begin{equation}
     e^{2A(r)}=\frac{1}{2}\left(1+\sqrt{1-2\gamma^2}\cosh{(2r)} \right) \ ,
\end{equation}
where $|\gamma|\leq 1/\sqrt{2}$. The dilaton profile is 
\begin{equation}
    \phi(r)=\phi_0+\frac{1}{\sqrt{2}}\log{\left( \frac{1+\sqrt{1-2\gamma^2}+\sqrt{2}\gamma\tanh{r}}{1+\sqrt{1-2\gamma^2}-\sqrt{2}\gamma\tanh{r}}\right)} \ .
\end{equation}
When $\gamma=0$, the dilaton becomes constant, proportional to $\phi_0$ and the metric reduces to pure $AdS_3$. When $\gamma=1/\sqrt{2}$, the spacetime is given by $\mathbb{R}\times AdS_2$. The minimal value of the warp factor in the Janus geometry is expressed as
\begin{equation}\label{eq::minimalwarpJanus}
    e^{A_*}=\left( \frac{1}{2}\left(1+\sqrt{1-2\gamma^2} \right)\right)^{1/2} \ .
\end{equation}

\subsection{Super Janus}\label{sec::reviewsuperJanus}

A Janus solution dual to a two-dimensional supersymmetric ICFT was originally constructed in \cite{Chiodaroli_2010}. This solution emerges within type IIB supergravity and it is asymptotically locally $AdS_3\times S^3\times M_4$, where the internal manifold $M_4$ can either be $T^4$ or $K_3$, reflecting the geometry of the target space in the dual CFT. For the sake of simplicity, we shall restrict our analysis to the case of $T^4$.

The ten-dimensional metric is constructed as a nontrivial fibration of $AdS_2\times S^2\times T^4$ over a Riemann surface $\Sigma$ and given by
\begin{equation}\label{eq::10dsuperJanus}
    ds^2_{10} = f_1^2 ds^2_{AdS_2} + f_2^2 ds^2_{S^2} + f_3^2 ds^2_{T^4} + \rho^2 ds^2_{\Sigma} \ ,
\end{equation}
where the warp factors $f_1^2, f^2_2, f_3^2$ and $\rho$ are nontrivial functions depending on coordinates along $\Sigma$ \cite{Chiodaroli_20102}. The solution is governed by a set of five constant parameters $\theta, \psi, L, k$ and $b$. After compactifying on $T^4$, the parameters $\psi$ and $\theta$ encode respectively the jump of the six-dimensional dilaton and axion across the interface. When $\theta=0$ and $\psi=0$ there is no interface and the spacetime becomes $AdS_3\times S^3\times T^4$ globally.

The compactification to three dimensions of \eqref{eq::10dsuperJanus} was derived in \cite{baig2024transmissioncoefficientsuperjanussolution} and the resulting metric takes the form \eqref{eq::defectmetric}, 
where the $AdS_3$ radius and the warp factor are 
\begin{equation}
    R^2=2L\cosh{\psi}\cosh{\theta}\ , \quad\quad e^{A(r)}=\frac{\cosh{r}}{\cosh{\psi}\cosh{\theta}} \ .
\end{equation}
In contrast to the RS braneworld, the minimal value of the warp factor in the super Janus background occurs at $r=0$ and it is given by
\begin{equation}
     e^{A_*} = \frac{1}{\cosh \psi \cosh \theta} \ .
\end{equation}

\section{Holographic Entanglement Entropy in ICFTs}\label{eq::secEEcomputations}

The entanglement entropy of a spatial region can be computed applying the Ryu-Takayanagi prescription \cite{Ryu_2006}, such that the EE is proportional to the area of a minimal surface in the dual geometry anchored at the boundary. For a $1+1$ field theory this is just the length of geodesics ending at the interval endpoints on the boundary.

\subsection{Minimal length geodesics}\label{eq::minlengthgeo}

In the presence of an interface the interval can be determined by the position of the endpoints of the interval relative to the position of the interface. We will denote by $l_L$ and $l_R$ the distance to the interface of the leftmost and rightmost endpoints of the interval respectively. The rightmost point will be taken always to be to the right of the interface, while the leftmost point could be to the left of the interface or to the right. We will denote as crossing interval the former case and non-crossing interval the latter, considering whether the associated geodesic connects or not the space at each side of the interface. For a BCFT ``crossing'' intervals will be those that have the leftmost point at the boundary, and will be characterized just by the total length $l=l_R$. In this case the geodesic ends at the boundary of space in the bulk of the dual geometry.

In the coordinates \eqref{eq::defectmetric}, the profile of the geodesic can be described by a function $x(r)$ obeying appropriate boundary conditions that depend on the interval. The profile is found by extremizing the following functional with respect to this function
\begin{equation}\label{eq:SEE}
    S=\frac{1}{4G}\int_{r_L}^{r_R} dr\, \mathcal{L},  
\end{equation}
where we have defined the Lagrangian
\begin{equation}\label{eq::Langrangian}
\mathcal{L}=R \,\sqrt{1+e^{2A(r)}\left( \frac{x'}{x}\right)^2} \ .
\end{equation}
Evaluated on the extremum, the value of $S$ equals the value of the entanglement entropy of the corresponding interval. This actually requires some regularization, since $S$ is divergent for geodesics ending at the boundary. We will use a cutoff in the $r$ coordinate to perform this regularization.

In order to find the solution for the profile we will take advantage of the scale isometry of  $AdS_2$ slices, $x\rightarrow \lambda x$. The Langrangian \eqref{eq::Langrangian}  is invariant under this transformation, and the associated Noether charge $c_s$ is given by
\begin{equation}\label{eq::cseq}
    c_s= \frac{\partial \mathcal{L}}{\partial x'}\frac{\partial (\lambda x)}{\partial \lambda}=\frac{R \,e^{2A}x'}{\sqrt{x^2+e^{2A}(x')^2}} \ .
\end{equation}
Solving for $x'$, \eqref{eq::cseq} yields
\begin{equation}\label{eq::diffeq}
    \frac{x'}{x}=\pm \frac{|c_s| e^{-A}}{\sqrt{ R^2 \,e^{2A}-c_s^2}} \ .
\end{equation}
We can take $c_s\geq 0$ in the expression above without loss of generality, with the sign corresponding to two possible branches for the solution. The type of profile depends on the value of $c_s$ relative to the value of the minimal warp factor $e^{A_*}$. If $c_s< R e^{A_*}$ the argument inside the square root \eqref{eq::diffeq} remains positive and the solution has no turning points in the $r$ coordinate, so that each branch corresponds to a complete solution. On the other hand, if $c_s> R e^{A_*}$ the solution has a turning point at the value of $r=r_{\text{min}}$ where the argument of the square root vanishes and the complete solution is found by the union of the two branches. The full geodesic then spans from infinity to $r_{\text{min}}$ in the radial coordinate.

It is important to note that $c_s$ is a non-trivial function of the ratio $l_L/l_R$ and its value alters the lengths of the intervals around or away from the interface, thereby affecting the EE. The ratio $l_L/l_R$ can be computed integrating \eqref{eq::diffeq}. In the rest of the paper we are going to limit ourselves to the case of $l_R\geq l_L$. However, we will often employ the left/right reflection symmetry $r\to -r$ of the background geometry. For solutions with no turning points, if we select the plus branch
\begin{equation}\label{eq::logratioequation}
    \log\left( \frac{l_L}{l_R}\right)=-\int_{-\infty}^{\infty}\frac{c_s e^{_-A}}{\sqrt{e^{2A}R^2-c_s^2}}dr =-2\int_0^{\infty}\frac{c_s e^{_-A}}{\sqrt{e^{2A}R^2-c_s^2}}dr \ .
\end{equation}
We will make use of \eqref{eq::logratioequation} repeatedly in the following section. For solutions with turning points 
\begin{equation}\label{eq::logratioequation2}
    \log\left( \frac{l_L}{l_R}\right)=-2\int_{r_{\text{min}}}^\infty\frac{c_s e^{_-A}}{\sqrt{e^{2A}R^2-c_s^2}}dr \ .
\end{equation}
Note that this last formula is valid for crossing intervals if we set $r_{\text{min}}=0$.

\begin{figure}[ht!] 
    \centering
    \includegraphics[width=0.75\textwidth]{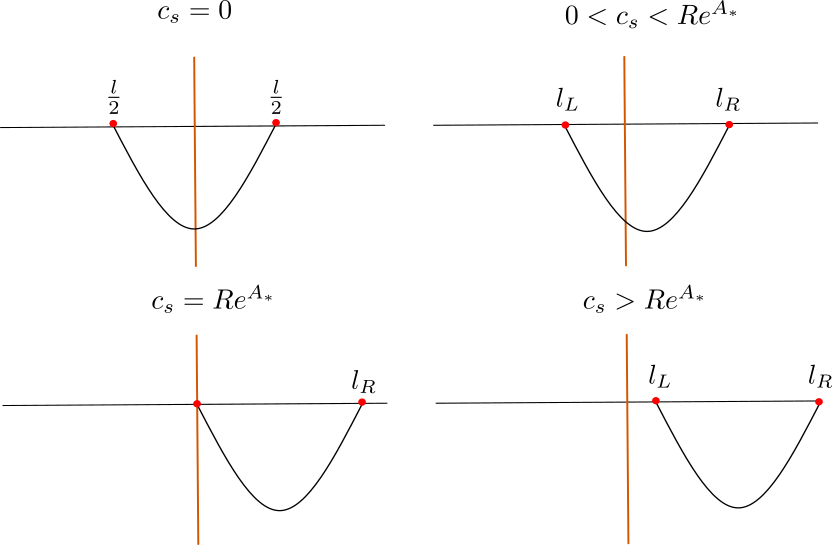}
    \caption{{We qualitatively depict here the four cases described above, where the orange vertical line is the interface at $r=0$ and the right/left half-lines represent the boundaries at $r=\pm\infty$. Going from left to right and top to bottom, we depict the fully symmetric case $c_s=0$, the generic interface crossing case $0<c_s< R\, e^{A_*}$, the fully asymmetric case $c_s= e^{A_*}$ and finally the non-interface crossing case $c_s>e^{A_*}$ .}}
    \label{fig:interface cases}
\end{figure}

\ \\
The type of profile for each of the cases is illustrated in figure \ref{fig:interface cases}:
\begin{itemize}
    \item $c_s=0$ yields $x'=0$ and thus $l_R=l_L=l/2$, which corresponds to crossing intervals symmetric around the interface. 
    
    \item $0<c_s<R \,e^{A_*}$ corresponds to the case of a crossing interval, with $l_R\geq l_L$, when choosing the plus sign in \eqref{eq::diffeq}, or $l_L\geq l_R$, when choosing the minus sign. This is the standard ICFT setup, discussed in \cite{Afxonidis:2024gne}. 
        
    \item $c_s=R\, e^{A_*}$ coincides with the case of having one endpoint interval on the interface, i.e. $l_L=0$ or $l_R=0$. In this case one of the endpoints reaches the boundary at $r=r_*$, i.e. $x(r_*)=0$. In this case one has to replace the lower limit of the integral in \eqref{eq::logratioequation} by $r_{\text{min}}=r_*$.
    This case will play an important role in the connection between the interface entropy and the effective central charge, as studied in Section \ref{sec::connectionloggceff}.
    
    \item $c_s=R e^{A(r_{\text{min}})}>R\, e^{A_*}$ corresponds to the case of non-crossing intervals with $r_{\text{min}}>r_*$, either on the left or the right of the interface. As we will see in Section \ref{sec::nics}, the EE associated to these geodesics acquires a finite contribution in the presence of the interface which is essential to ensure strong subadditivity, as discussed in Section \ref{sec::validitySSA}.
\end{itemize}

\subsection{Entanglement entropy}\label{sec::EE}

The EE associated to each interval is computed by evaluating the Lagrangian on the profile and integrating over the $r$ direction. The on-shell Lagrangian is
\begin{equation}\label{eq::onshellLang}
    \mathcal{L}=\frac{R}{\sqrt{1-c_s^2 e^{-2A} R^{-2}}} \ .
\end{equation}
When $r\to \pm\infty$, the warp factor approaches the $AdS$ value
\begin{equation}\label{eq:warpAppr}
    e^A\approx \frac{1}{2} e^{a_\pm}e^{\pm r} \to +\infty,
\end{equation}
with $a_\pm$ constant. Therefore the on-shell Lagrangian approaches a constant $\mathcal{L}\approx R$ and the entanglement entropy diverges when the limits in \eqref{eq:SEE} are taken to infinity. In order to get a finite result we need to regularize the integral, which we do by introducing cutoffs $r_L,r_R$ in the radial direction. The regularized EE for a crossing interval is, taking into account \eqref{eq:centralc},
\begin{equation}\label{eq:SAc}
    S_A=\frac{c}{6}\int_{-r_L}^{r_R}\frac{dr}{\sqrt{1-c_s^2 e^{-2A} R^{-2}}} \ .
\end{equation}
For a non-crossing interval on the right side of the interface we have instead
\begin{equation}\label{eq:SAnc}
    S_A=\frac{c}{6}\left(\int_{r_{\text{min}}}^{r_L}+\int_{r_{\text{min}}}^{r_R}\right)\frac{dr}{\sqrt{1-c_s^2 e^{-2A} R^{-2}}} \ ,
\end{equation}
If there is a reflection symmetry $r\to -r$ we could in principle use \eqref{eq:SAnc} for a crossing interval setting the lower limit to $r_{\text{min}}=0$. The regularized entropy can also be constructed by separating the finite and divergent pieces. Then, assuming there is reflection symmetry $r\to -r$, the EE can be written for both crossing and non-crossing intervals as
\begin{equation}\label{eq:SA}
    S_A=\frac{c}{6}\left[2\int_{r_\text{min}}^{\infty}\left( \frac{1}{\sqrt{1-c_s^2 e^{-2A} R^{-2}}}-1 \right)+r_L+r_R-2r_\text{min}\right] \ .
\end{equation}
We will now see how to connect the radial cutoff to the cutoff in field theory. Consider for simplicity a symmetric crossing interval of length $l$, where $r_L=r_R=r_c$ and $S_A=\frac{cr_c}{3}$. In the absence of an interface the geometry is $AdS$. Choosing a position independent cutoff procedure when the metric on the 2d CFT spacetime is the standard Minkowski one, the cutoff in Poincar\'e coordinates \eqref{eq:poincarecoord} lies at $z=\epsilon$, in such a way that, close to the rightmost endpoint of the interval
\begin{equation}
    \epsilon=\frac{l}{2\cosh r_c}\approx l e^{-r_c}.
\end{equation}
Solving for $r_c$ we recover the usual result for the EE in a CFT
\begin{equation}
    r_c=\log\frac{l}{\epsilon}\ \Rightarrow \ S_A^{  ^{\rm CFT}}=\frac{c}{3}\log\frac{l}{\epsilon}.
\end{equation}
The generalization to other spaces and intervals is, for the rigthmost endpoint,
\begin{equation}
    \epsilon= l_R e^{-A(r_R)} \approx 2l_R e^{-a_+}e^{- r_R}.
\end{equation}
Therefore
\begin{equation}\label{eq:cutoffR}
    r_R=-a_++\log\frac{2l_R}{\epsilon}.
\end{equation}
The $r_L$ cutoff depends on whether the interval is crossing or non-crossing
\begin{equation}\label{eq:cutoffL}
    \text{crossing:}\ \ r_L=-a_-+\log\frac{2l_L}{\epsilon},\quad \text{non-crossing:}\ \ r_L=-a_++\log\frac{2l_L}{\epsilon}.
\end{equation}
If the space has reflection symmetry in $r$, then $a_+=a_-$ and there is no difference between the cutoff procedure for the crossing and non-crossing intervals.

For the symmetric crossing interval the EE in the general case becomes
\begin{equation}
    S_A^{ ^{\rm ICFT}}=\frac{c}{3}\log\frac{l}{\epsilon}-\frac{c}{6}(a_++a_-).
\end{equation}
Therefore, there is an additional finite contribution to the EE which can be identified with the boundary entropy number in a symmetric interval
\begin{equation}\label{eq:gisym}
    \log g=-\frac{c}{6}(a_++a_-).
\end{equation}
In the following, we will assume reflection symmetry $a_+=a_-=a$, thus $\log g=-\frac{ca}{3}$.

Although we do expect to find a non-zero contribution compared to a CFT, it should be noted that this result is scheme-dependent, multiplying the cutoff $\epsilon$ by a factor (which is equivalent to shifting the coordinate $r$) would shift the value of the finite contribution to the EE. On the other hand, the difference between the EE of intervals of the same length and other similar quantities, such as the mutual information of two disjoint intervals, should be scheme independent. These quantities can capture contributions from the interface that are absent in the CFT. We will provide later a scheme-independent definition of interface entropy using asymmetric intervals.

\subsection{Interface entropy for crossing intervals}\label{sec::ics}

We are now going to focus on crossing intervals, with endpoints at distances $l_L$ and $l_R$ from the interface. As mentioned before we will take $l_R\geq l_L$.

In the reflection symmetric case the EE is given by \eqref{eq:SA} with $r_{\text{min}}=0$. The dependence on the cutoffs will be similar in the general case, but the integral should be taken as in \eqref{eq:SAc}. Taking into account \eqref{eq:cutoffR} and \eqref{eq:cutoffL}, the EE for a crossing interval $A$ can be expressed as \cite{Afxonidis:2024gne,Kruthoff:2021vgv}
\begin{equation}\label{eq::EEicc}
    S_A=\frac{c}{6}\log{\left( \frac{2l_L }{\epsilon}\right)}+\frac{c}{6}\log{\left( \frac{2l_R }{\epsilon}\right)}+\log g^{(2)} \ ,
\end{equation}
Where we identify the interface entropy as
\begin{equation} \label{eq::loggicc}
    \log g^{(2)} =\frac{c}{6}\int_{-\infty}^{\infty}dr\left(\frac{1}{\sqrt{1-c_s^2e^{-2A}R^{-2}}}-1 \right)+\log g\ .
\end{equation}
Note that this is not the standard way to define the interface entropy. Usually the EE is given as
\begin{equation}\label{eq::EEicc2}
    S_A=\frac{c}{3}\log{\left( \frac{l_L+l_R }{\epsilon}\right)}+\log g^{(1)} \ ,
\end{equation}
with $\log g^{(1)}$ the interface entropy. The two differ by a finite term depending on the position of the endpoints of the interval
\begin{equation}\label{eq::g1g2relation}
    \log g^{(1)}= \log g^{(2)} + \frac{c}{6}\log \frac{4l_L/l_R}{(1+l_L/l_R)^2} \ .
\end{equation}
In \eqref{eq::EEicc2}, the EE takes the same form as the one of an interval of length $l_L+l_R$ in a CFT plus an additional piece $\log g^{(1)}$ that would vanish in the absence of the interface. On the other hand, in \eqref{eq::EEicc}, the EE takes the same form as the sum of EEs of two independent BCFTs, each at one side of the interface, with intervals of length $l_L$ and $l_R$ ending at their boundaries. The finite term $\log g^{(2)}$ would capture the sum of the boundary entropies of each BCFT plus a possible additional contribution if the two BCFTs are not completely decoupled. For example, if we have no interface $\log g^{(1)}$=0, but $\log g^{(2)}\neq 0$, since this is equivalent to putting an effective fully transmissive interface that couples the BCFTs. Note that regardless of the form we choose, the value of $S_A$ is the same. For a symmetric crossing interval $l_L=l_R=l/2$ there is no difference between the two ways of expressing the EE and $\log g^{(1)}=\log g^{(2)}=\log g$, given in \eqref{eq:gisym}.

\subsection{Interface entropy for non-crossing intervals}\label{sec::nics}

For non-crossing intervals with $l_R>l_L>0$, the EE is again given by \eqref{eq::EEicc}, but with an interface entropy
\begin{equation} \label{eq::loggicc2}
    \log g^{(2)} =\frac{c}{3}\left[\int_{r_{\text{min}}}^{\infty}dr\left(\frac{1}{\sqrt{1-c_s^2e^{-2A}R^{-2}}}-1 \right)-r_{\text{min}}\right]+\log g\ .
\end{equation}
In this case the usual form of the EE is, for an interval of length $l_R-l_L$,
\begin{equation}\label{eq::EEicc3}
    S_A=\frac{c}{3}\log{\left( \frac{l_R-l_L }{\epsilon}\right)}+\log g^{(1)} \ .
\end{equation}
This changes the relation between the two types of interface entropy
\begin{equation}\label{eq::g1g2relationnicc}
    \log g^{(1)} = \log g^{(2)} + \frac{c}{6}\log \frac{4l_L/l_R}{(1-l_L/l_R)^2} \ .
\end{equation}
An interesting observation is that there is a non-zero finite contribution to the interface entropy $\log g^{(1)}$ when $l_L=l_R$ and the length of the interval goes to zero. For $l_R-l_L\to 0$ we have $c_s\to \infty$ and $r_{\text{min}}\to \infty$. The warp factor in \eqref{eq::logratioequation2} and \eqref{eq::loggicc2} can be approximated by \eqref{eq:warpAppr}, and one finds to leading order
\begin{equation}
    \log g^{(2)}\approx \frac{c}{3} (\log 2-r_{\text{min}})+\log g,\qquad \log\left(\frac{l_L}{l_R}\right)\approx -4e^{-a}e^{-r_{\text{min}}}\ .
\end{equation}
Combining these with \eqref{eq::g1g2relationnicc}, we arrive at
\begin{equation}
    \log g^{(1)}\approx \frac{c}{3}\log\left[-\frac{1}{2}e^a\log\left(\frac{l_L}{l_R}\right)\right]+\frac{c}{6}\log \frac{4l_L/l_R}{(1-l_L/l_R)^2}+\log g\, .
\end{equation}
Expanding in the length of the interval $l=l_R-l_L\to 0$, we arrive at
\begin{equation}\label{eq:finitelogg1zerolength}
\log g^{(1)} =\frac{ca}{3}+\log g= 0\ . 
\end{equation}
Thus far away from the interface one recovers the result of a CFT without interfaces.

\subsection{Monotonicity of interface entropy with the ratio $l_L/l_R$}

The interface entropy $\log g^{(2)}$ is a monotonically decreasing function of the ratio  $l_L/l_R$ as stated by the $g_{\text{eff}}$-theorem discussed in \cite{Afxonidis:2024gne}.

Let us define the ratio between the distance of the endpoints to the interface as $\rho=l_L/l_R$ \footnote{Not to be confused with the warp factor in the super Janus metric.}, and the associated pseudo-beta function for the interface entropy
\begin{equation}\label{eq:Bg}
    \BB=\frac{d\log g^{(2)}}{d \log \rho}\ .
\end{equation}
As explained in \cite{Afxonidis:2024gne} this pseudo-beta function, which describes the evolution of the EE as the function of a conformally invariant cross ratio of length scales, is not an actual beta function describing evolution of the EE under RG flows, but instead simply encodes its dependence on a particular geometric quantity within the ICFT. If $\BB$ has a definite sign, then the interface entropy will be a monotonic function of the ratio, either increasing for $\BB> 0$, or decreasing  for $\BB<0$. In order to find its value we use $c_s$ as the varying parameter, so that
\begin{equation}
    \BB=\frac{d\log g^{(2)}}{d \log \rho}=\frac{d\log g^{(2)}}{dc_s}\frac{dc_s}{d\log \rho}\ .
\end{equation}
Now, using the expressions in \eqref{eq::logratioequation} and \eqref{eq::loggicc}, we can see that 
\begin{equation}\label{eq::firstderivative}
  \frac{d \log \rho}{d c_s}=-\int_{-\infty}^{\infty}\frac{e^AR^2}{\left( e^{2A}R^2-c_s^2\right)^{3/2}}<0\ ,\quad \quad \frac{d\log g^{(2)}}{d c_s}=\frac{c}{6} \int_{-\infty}^{\infty}\frac{c_s e^AR}{\left( e^{2A}R^2-c_s^2\right)^{3/2}}>0  \ .
\end{equation}
It follows that $\log g^{(2)}$ is a monotonically decreasing function of the ratio for crossing intervals
\begin{equation}\label{eq:Bgc}
    \BB^c=-\frac{c}{6}\frac{c_s}{R}<0 \ .
\end{equation}
For a non-crossing interval we can do a similar analysis, but using $r_{\text{min}}$ as a varying parameter. Taking the expressions in \eqref{eq::logratioequation2} and \eqref{eq::loggicc2}, we find
\begin{equation}\label{eq::firstderivative2}
  \frac{d \log \rho}{d r_{\text{min}}}\sim 2\frac{c_s e^{-A(r_{\text{min}})}}{\sqrt{e^{2A(r_{\text{min}})}R^2-c_s^2}}>0\ ,\quad \quad \frac{d\log g^{(2)}}{d r_{\text{min}}}\sim -\frac{c}{3}\frac{e^{A(r_{\text{min}})}R}{\sqrt{e^{2A(r_{\text{min}})}R^2-c_s^2}} <0  \ .
\end{equation}
This gives
\begin{equation}\label{eq:Bgnc}
    \BB^{nc}=-\frac{c}{6}\frac{e^{2A(r_{\text{min}})}R}{c_s}=-\frac{c}{6}\frac{c_s}{R}<0 \ .
\end{equation}
In this case $\log g^{(2)}$ is also a monotonically decreasing function of the ratio.

\section{Monotonicity of the interface entropy and Strong Subadditivity}\label{sec::validitySSA}

Strong Subadditivity (SSA) is one of the main properties the EE of a unitary theory has to satisfy. In \cite{Afxonidis:2024gne} SSA was used to prove the entropic $g_{\text{eff}}$-theorem for crossing intervals. We can apply a similar argument to derive a $g_{\text{eff}}$-theorem for non-crossing intervals.

\begin{figure}[h!] 
    \centering
    \includegraphics[width=0.6\textwidth]{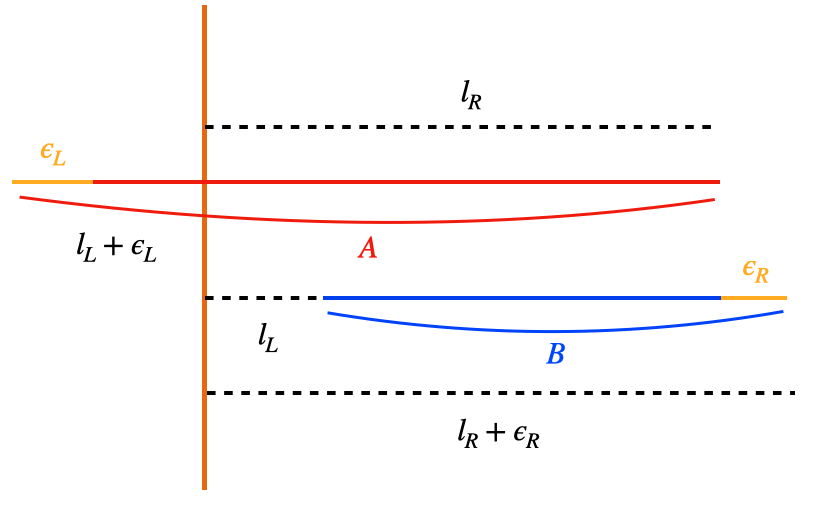}
    \caption{Setup for SSA for non-crossing intervals. The orange line represents the interface.}
    \label{fig:SSAcrossnoncross}
\end{figure}

We can express the EE for an interval with arbitrary $l_L$ and $l_R$ as 
\begin{equation}\label{eq::EEwithF}
     S(l_L, l_R) = \frac{c}{6} \log\frac{2l_L}{\epsilon}+\frac{c}{6} \log\frac{2l_R}{\epsilon} + F\left( \frac{l_L}{l_R} \right) \ , \quad F\left( \frac{l_L}{l_R}\right)\equiv\log g^{(2)}\left( \frac{l_L}{l_R}\right) \ ,
\end{equation}
The function $F$  depends on whether the interface is inside the interval or not, so we will denote it as $F_c$ for crossing intervals, and $F_{nc}$ for non-crossing intervals.
To apply the SSA condition, we first consider the configuration shown in Figure \ref{fig:SSAcrossnoncross}, where  $\epsilon_L>0$ and $\epsilon_R>0$.
We can thus express the SSA as follows
\begin{equation}
    S_c\left(l_L+\epsilon_L,l_R\right)+S_{nc}\left(l_L,l_R+\epsilon_R\right)-S_{c}\left(l_L+\epsilon_L,l_R+\epsilon_R\right)-S_{nc}\left(l_L,l_R\right) \geq 0 \ .
\end{equation}
Expanding for infinitesimally small $\epsilon_L$ and $\epsilon_R$ we obtain for the leading order contribution
\begin{equation}
    -\epsilon_R\left(\frac{\partial S_{c}(l_L,l_R)}{\partial l_R}-\frac{\partial S_{nc}(l_L,l_R)}{\partial l_R} \right) + \mathcal{O}\left( \epsilon^2\right) \geq 0 \ .
\end{equation}
Since $\epsilon_R>0$ we arrive at
\begin{equation}\label{eq::SSAcrossnoncrosscond}
    \frac{\partial S_{c}(l_L,l_R)}{\partial l_R}-\frac{\partial S_{nc}(l_L,l_R)}{\partial l_R} \leq 0 \ .
\end{equation}
Plugging in \eqref{eq::EEwithF} into \eqref{eq::SSAcrossnoncrosscond} we obtain
\begin{equation}\label{eq::c0c0tildecond}
   \rho F_{nc}'(\rho)\leq \rho F_{c}'(\rho) \leq 0 \ ,
\end{equation}
where $\rho = l_L/l_R$ and we have employed in the second inequality the $g_{\text{eff}}$-theorem \cite{Afxonidis:2024gne}. Equation \eqref{eq::c0c0tildecond} implies that the interface entropy of non-crossing intervals exhibits a similar monotonically decreasing behavior as for crossing intervals 
\begin{equation}
    \frac{d\log g^{(2)}(\rho)}{d\rho} \leq 0 \ .
\end{equation}
In addition, taking the limit when $\rho\to 0$ in \eqref{eq::c0c0tildecond} and defining $\lim_{\rho \to 0}(\rho F'_c(\rho))=c_0$ and $\lim_{\rho \to 0}(\rho F'_{nc}(\rho))=\tilde{c}_0$ yields
\begin{equation}
    -c_0\leq -\tilde{c}_0 \leq 0 \ .
\end{equation}
More generally, \eqref{eq::c0c0tildecond} indicates that the rate of change of $\log g^{(2)}$ for non-crossing intervals must be faster or equal than for crossing intervals.  Therefore the SSA indicates that finite EE contributions in non-crossing intervals must be present whenever $\log g^{(2)}$ changes with the ratio, while the na\"{\i}ve expectation would have been that only crossing intervals receive a finite contribution to the EE from the interface. This is one of our main results.

\begin{figure}[ht!] 
    \centering
    \includegraphics[width=0.6\textwidth]{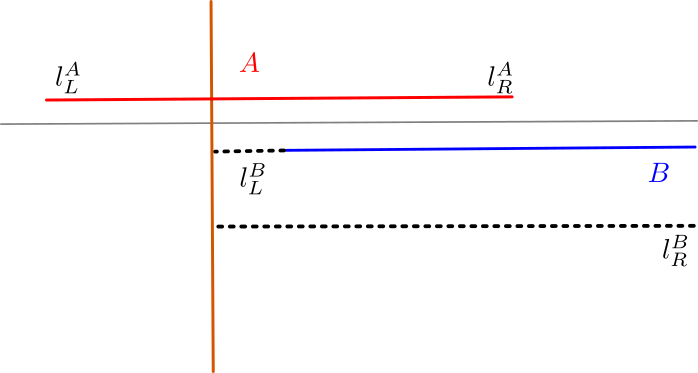}
    \caption{Setup for SSA. The orange line represents the interface.}
    \label{fig:SSAtimeslice}
\end{figure}

Let us now confirm these conclusions in  holographic ICFTs. We can do this for generic intervals $A$ and $B$ arranged as in Figure \ref{fig:SSAtimeslice}, characterized by the distance of the endpoints to the interface. 

The property of SSA establishes that 
\begin{equation}\label{eq:SSAcond}
    S_A+S_B-S_{A\cup B}-S_{A\cap B}\geq 0 \ .
\end{equation}
 Using the expression in \eqref{eq::EEicc}, the cutoff-dependent terms cancel out, since $l_L^{A\cap B}=l_L^B$, $l_R^{A\cap B}=l_R^A$, $l_L^{A\cup B}=l_L^A$ and $l_R^{A\cup B}=l_R^B$. Therefore we arrive at the condition
\begin{equation}\label{eq:SSAcondlogg}
    \log g^{(2)}_A+\log g^{(2)}_B-\log g^{(2)}_{A\cup B}-\log g^{(2)}_{A\cap B} \geq 0\,.
\end{equation}
We have already seen that in holographic ICFTs $\log g^{(2)}$ for both crossing and non-crossing intervals is a monotonically decreasing function of the ratio $l_L/l_R$. Since in general $l_L^A/l_R^A \geq  l_L^{A\cup B}/l_R^{A\cup B}$, it follows that crossing intervals contribute negatively to the SSA condition
\begin{equation}
    \log g^{(2)}_A-\log g^{(2)}_{A\cup B}\leq 0.
\end{equation}
On the other hand, $l_L^B/l_R^B \leq  l_L^{A\cap B}/l_R^{A\cap B}$, so non-crossing intervals contribute positively to the SSA condition
\begin{equation}
    \log g^{(2)}_B-\log g^{(2)}_{A\cap B} \geq 0\,.
\end{equation}
Therefore, as anticipated, in order for the SSA condition to be satisfied, not only $\log g^{(2)}$ of non-crossing intervals has to be monotonically decreasing with $l_L/l_R$, they also must do so faster than for crossing intervals.

Let us define the ratios
\begin{equation}
    \rho_A=\frac{l_L^A}{l_R^A}\ , \quad \rho_{A\cup B}=\frac{l_L^A}{l_R^B}\ , \quad \rho_B=\frac{l_L^B}{l_R^B}\ , \quad \rho_{A\cap B}=\frac{l_L^B}{l_R^A} \ .
\end{equation}
The SSA condition is then
\begin{equation}
    \log g^{(2)}(\rho_A)+ \log g^{(2)}(\rho_B)- \log g^{(2)}(\rho_{A\cup B})- \log g^{(2)}(\rho_{A\cap B}) \geq 0 \ .
\end{equation}
Given the pseudo-beta functions $\BB$ defined in \eqref{eq:Bg}, we rewrite the above equation as
\begin{equation}
    \int_{\rho_{A \cup B}}^{\rho_A} \frac{d\rho}{\rho}\, \BB^c- \int^{\rho_{A \cap B}}_{\rho_B} \frac{d\rho}{\rho}\, \BB^{nc}\geq 0 \ .
\end{equation}
Where $\BB^c$ refers to crossing intervals and $\BB^{nc}$ to non-crossing intervals. Now using \eqref{eq:Bgc} and \eqref{eq:Bgnc}, we rearrange the SSA condition as
\begin{equation}
    \int^{\rho_{A \cap B}}_{\rho_B} \frac{d\rho}{\rho} \,c_s^{nc}(\rho) \geq \int_{\rho_{A \cup B}}^{\rho_A} \frac{d\rho}{\rho}\, c_s^c(\rho) \ ,
\end{equation}
where we make explicit the fact that $c_s$ is a function of the ratio. We will now prove the above bound. First note that
\begin{equation}
 \int^{\rho_{A \cap B}}_{\rho_B} \frac{d\rho}{\rho} \,c_s^{nc}(\rho) \geq     \left({\min_{\rho \in [\rho_{A \cap B},\rho_B]} c_s^{nc} (\rho)}\right)\int^{\rho_{A \cap B}}_{\rho_B} \frac{d\rho}{\rho}\ ,
\end{equation}
 and
 \begin{equation}
  \left(\max_{\rho \in [\rho_A,\rho_{A \cup B}]} c_s^c(\rho) \right)\int_{\rho_{A \cup B}}^{\rho_A} \frac{d\rho}{\rho} \geq \int_{\rho_{A \cup B}}^{\rho_A} \frac{d\rho}{\rho}\, c_s^c(\rho) \ .
 \end{equation}
On the other hand 
\begin{equation}
     \int^{\rho_{A \cap B}}_{\rho_B} \frac{d\rho}{\rho} = \log{\frac{\rho_{A \cap B}}{{\rho_B}}}=\log\frac{l_R^B}{l_R^A}=\log{\frac{\rho_A}{\rho_{A \cup B}}}=\int_{\rho_{A \cup B}}^{\rho_A} \frac{d\rho}{\rho} \ .
\end{equation}
Therefore, the SSA condition follows from the fact that
\begin{equation}
    {\min_{\rho \in [\rho_{A \cap B},\rho_B]} c_s^{nc} (\rho)}\geq R e^{A_*} \geq \max_{\rho \in [\rho_A,\rho_{A \cup B}]} c_s^c(\rho)\geq 0 \ ,
\end{equation}
which is a consequence of $B_g<0$ and $|\BB^{nc}| > |\BB^c|$ .

\section{Interface entropy and effective central charge}\label{sec::connectionloggceff}

We will present in this section another  main result, the connection between the interface entropy $\log g^{(2)}$ and the effective central charge $c_{\text{eff}}$ that arises when an endpoint of the interval is at the interface. Let us consider $l_L=0$ (so $l_R=l$), then the EE becomes \cite{karch2023universalityeffectivecentralcharge,Sakai:2008tt,Brehm:2015lja}
\begin{equation}\label{eq:SAceff}
    S_A=\frac{c+c_{\text{eff}}}{6}\log\frac{l}{\epsilon}+\log \tilde{g}\ ,
\end{equation}
with $c_{\text{eff}}\leq c$. In \cite{Gutperle_2016,karch2023universalityeffectivecentralcharge} it was found that, for holographic ICFTs,
\begin{equation}\label{eq:ceff}
    c_{\text{eff}}=\frac{3Re^{A_*}}{2G}=c \,e^{A_*},
\end{equation}
with $e^{A_*}\leq 1$ the minimal value of the warp factor. The same expression for $c_{\text{eff}}$ has been derived in the context of holographic RG interfaces as well \cite{Gutperle:2024yiz}. Since for $l_L>0$ the coefficient of the logarithmic term is $c/3$, this looks like a discontinuous jump in the coefficient of the universal term. However, as we explain below, the change can be understood as a consequence of the growth of the interface entropy in the limit $l_L/l_R\rightarrow 0$. 

The first observation is that introducing a cutoff to regularize the EE implies that the location of the interface can only be determined up to distances of the order of the cutoff. For some value of $l_L\approx \lambda \epsilon$ the minimal length geodesic will arrive at the cutoff at $r_L\approx r_\epsilon\equiv -a_-+\log(2\lambda)$ for a crossing interval or $r_L\approx r_*$ for a non-crossing one. In this case, taking $l_R\approx l$, and introducing \eqref{eq::g1g2relation} in \eqref{eq::EEicc2}, or \eqref{eq::g1g2relationnicc} in \eqref{eq::EEicc3}, we arrive at
\begin{equation}
    S_A\approx \frac{c}{6}\log\frac{l}{\epsilon}+\frac{c}{6}\log (4\lambda)+\log g^{(2)}\,.
\end{equation}
The non-trivial result that we will show in the following is that $\log g^{(2)}$ is logarithmically divergent with the right coefficient to account for the $c_{\text{eff}}$ contribution. We will also compute the remaining finite contribution to the entanglement entropy, and identify a scheme-independent definition of the interface entropy. We will evaluate these quantities for the examples introduced in Section \ref{sec::reviewJanussuperJanus}.

\subsection{The $l_L/l_R\to 0$ limit of the interface entropy}

The limit $l_L/l_R\to 0$ corresponds to the limit $c_s\to Re^{A_*}$ and, for non-crossing intervals, $r_{\text{min}}\to r_*$\ . In this limit the integrals \eqref{eq::logratioequation}, \eqref{eq::logratioequation2} and \eqref{eq::loggicc}, \eqref{eq::loggicc2}  become divergent. In order to see this, note that the integration includes or comes very close to the point $r_*$ where the warp factor reaches its minimal value,  
\begin{equation}
    e^{A(r)}\approx e^{A_*}+b_2 (r-r_*)^2+\dots \ ,
\end{equation}
where $b_2>0$ is a constant. For $c_s=Re^{A_*}$ the integrands close to $r_*$ have a factor
\begin{equation}
    \frac{1}{\sqrt{e^{2A}R^2-c_s^2}}\approx \frac{1}{R\sqrt{(e^{A_*}+b_2(r-r_*)^2)^2-e^{2A_*}}}\approx\frac{1}{R\sqrt{b_2}|r-r_*|}\,.
\end{equation}
Therefore, the integrals have the same type of logarithmic divergence when integrated around $r_*$.
Let us split the interface entropy and $\log(l_L/l_R)$ into a finite regularized part plus a divergent part
\begin{equation}\label{eq::splitreganddiv}
    \log g^{(2)}=\log g_{\text{reg}}^{(2)}+\log g_{\text{div}}^{(2)} \ , \quad\quad
    \log \left( \frac{l_L}{l_R}\right)=\log \left( \frac{l_L}{l_R}\right)_{\text{reg}}+\log \left( \frac{l_L}{l_R}\right)_{\text{div}} \ .
\end{equation}
The divergent parts are found by expanding the integrands of \eqref{eq::logratioequation}, \eqref{eq::logratioequation2} and \eqref{eq::loggicc}, \eqref{eq::loggicc2} close to $r_*$. If there is reflection symmetry, then for the crossing interval there is a second divergence at $r=-r_*$. We introduce an integral that reproduces the divergences. For a non-crossing interval it takes the form
\begin{equation}\label{eq::logglogratioregdiv0}
    \mathcal{I}=\int_{r_{\text{min}}}^{r_0}\frac{dr}{\sqrt{e^{2A_*}-c_s^2R^{-2}+2e^{2A_*}b_2(r-r_*)^2}} \ ,
\end{equation}
The upper limit is an arbitrary value $r_0> r_{\text{min}}$. Note that, in the $l_L/l_R\to 0$ limit
\begin{equation}
r_{\text{min}}\approx r_*+\sqrt{\frac{c_s^2R^{-2}-e^{2A_*}}{2 e^{2A_*}b_2}}    \,.
\end{equation}
For a crossing interval, if we assume reflection symmetry,
\begin{equation}\label{eq::logglogratioregdiv2}
    \mathcal{I}=\frac{1}{2}\int_0^{r_0}\frac{dr}{\sqrt{e^{2A_*}-c_s^2R^{-2}+2e^{2A_*}b_2(r-r_*)^2}}+\frac{1}{2}\int_{-r_\epsilon}^{0}\frac{dr}{\sqrt{e^{2A_*}-c_s^2R^{-2}+2e^{2A_*}b_2(r+r_*)^2}} \ .
\end{equation}
Where $r_0,r_\epsilon>r_*$. We can make the expression symmetric by fixing the position of the cutoff to
\begin{equation}
    r_\epsilon=r_0\ \Rightarrow \ \lambda=\frac{1}{2}e^{a_-}e^{r_0}\,.
\end{equation}
In this case, for both crossing and non-crossing intervals
\begin{equation}\label{eq::logglogratioregdiv}
    \mathcal{I}=\int_{r_{\text{rmin}}}^{r_0}\frac{dr}{\sqrt{e^{2A_*}-c_s^2R^{-2}+2e^{2A_*}b_2(r-r_*)^2}} \ ,
\end{equation}
with $r_{\text{rmin}}=0$ for  crossing intervals.

The divergent parts in both cases are
\begin{equation}\label{eq::logglogratioregdiv}
\log g_{\text{div}}^{(2)}=\frac{c\,e^{A_*}}{3}\mathcal{I},\qquad \log \left( \frac{l_L}{l_R}\right)_{\text{div}}=-\frac{2c_se^{-A_*} }{R}\mathcal{I}\,.
\end{equation}
One can compute analytically the integral ${\cal  I}$ and show that there is a divergence 
\begin{equation}
    {\cal I}\sim -\log\left|e^{2A_*}-c_s^2R^{-2} \right|.
\end{equation}
The finite regularized parts are defined subtracting these quantities from the full expressions \eqref{eq::logratioequation}, \eqref{eq::logratioequation2} and \eqref{eq::loggicc}, \eqref{eq::loggicc2}. Combining the two formulas in \eqref{eq::logglogratioregdiv}, we arrive at
\begin{equation}\label{eq:logg2complete}
\log g^{(2)}=-\frac{c e^{2A_*}R}{6c_s} \log \left( \frac{l_L}{l_R}\right)+\frac{c e^{2A_*}R}{6c_s}\log \left( \frac{l_L}{l_R}\right)_{\text{reg}}+\log g_{\text{reg}}^{(2)}\,.
\end{equation}
Then, when the endpoint of the interval is at a distance of the order of the cutoff from the interface, $l_L\approx \lambda \epsilon$, $l_R\approx l$ and $c_s\approx R e^{A_*}$ we find 
\begin{equation}\label{eq:logg2completeLimit}
\log g^{(2)}\approx \frac{c_{\text{eff}}}{6} \log \left( \frac{l}{\epsilon}\right)-\frac{c_{\text{eff}}}{6} \log \left(\lambda\right)+\frac{c_{\text{eff}}}{6} \log \left( \frac{l_L}{l_R}\right)_{\text{reg}}+\log g_{\text{reg}}^{(2)}\ .
\end{equation}
where we have used \eqref{eq:ceff}. The first term in the previous formula accounts for the logarithmic contribution proportional to $c_{\text{eff}}$ in \eqref{eq:SAceff}, as anticipated. Note that the coefficient of $\log(l_L/l_R)$ in  \eqref{eq:logg2complete} is inversely proportional to the pseudo-beta function \eqref{eq:Bgc},\eqref{eq:Bgnc}.

\subsection{Finite contributions in the $l_L/l_R\to 0$ limit}

The finite part of the EE in the $l_L/l_R\to 0$ limit is, after adding all possible terms
\begin{equation}
    \log \tilde g=\frac{c}{6}\log (4\lambda)-\frac{c_{\text{eff}}}{6} \log \left(\lambda\right)+\frac{c_{\text{eff}}}{6} \log \left( \frac{l_L}{l_R}\right)_{\text{reg}}+\log g_{\text{reg}}^{(2)}\,.
\end{equation}
Aside from the explicitly scheme-dependent part depending on $\lambda$, we will be interested in finding the value of the additional finite contributions.

\begin{figure}[H] 
    \centering
    \includegraphics[width=0.7\textwidth]{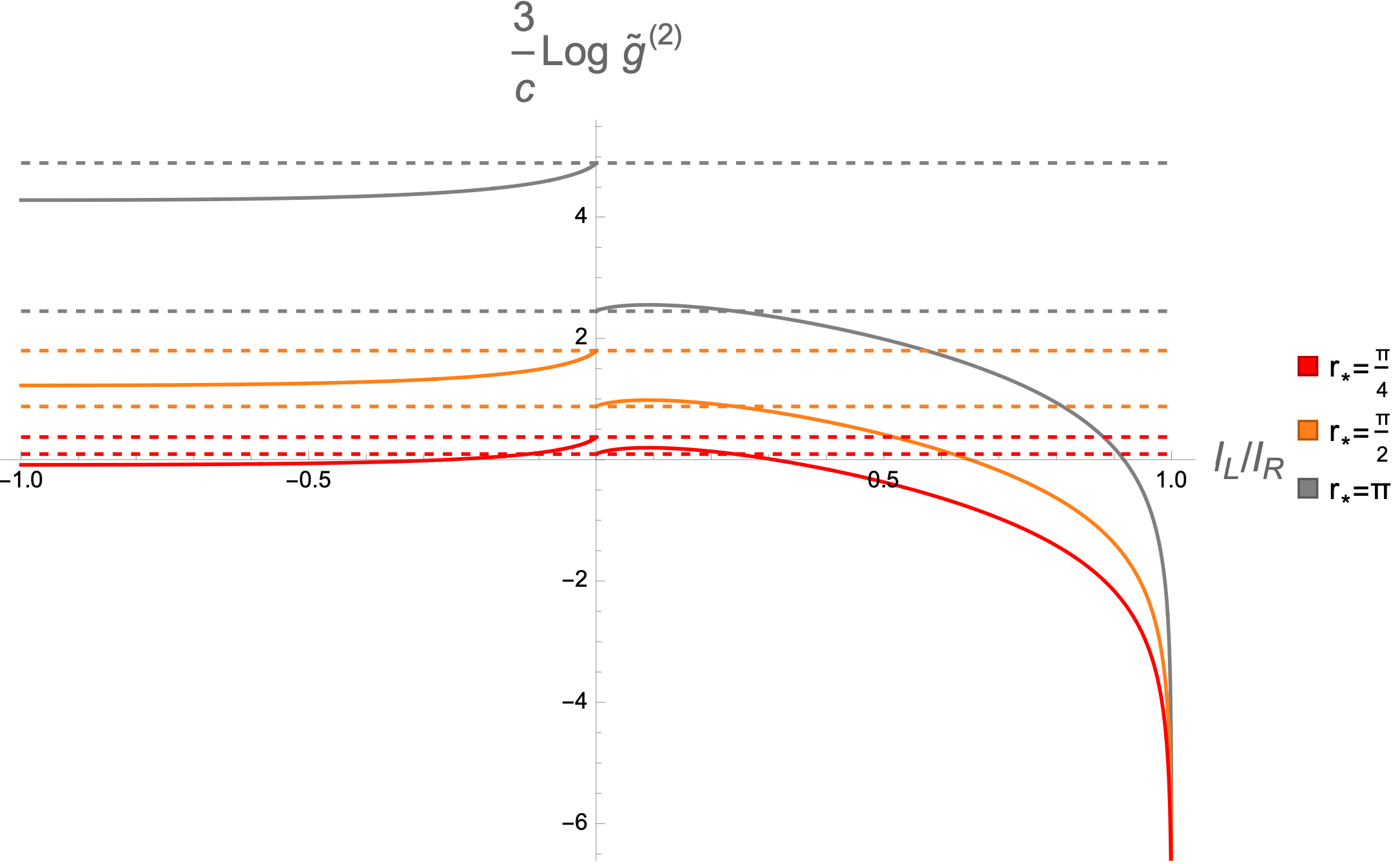}
    \caption{Plot of the scheme-independent contribution to the interface entropy \eqref{eq:logg2tilde} as a function of the ratio $ l_L/l_R$ for the RS braneworld geometry, for different values of the parameter $r_*$. Negative values correspond to crossing intervals and positive values to non-crossing intervals. The dashed lines correspond to the values in \eqref{eq:logg2tilde0}. Note that there is a jump between crossing and non-crossing intervals, given by \eqref{eq:RSbraneJump}.}
    \label{fig:RSlogg2}
\end{figure}

\begin{figure}[H] 
    \centering
\includegraphics[width=0.7\textwidth]{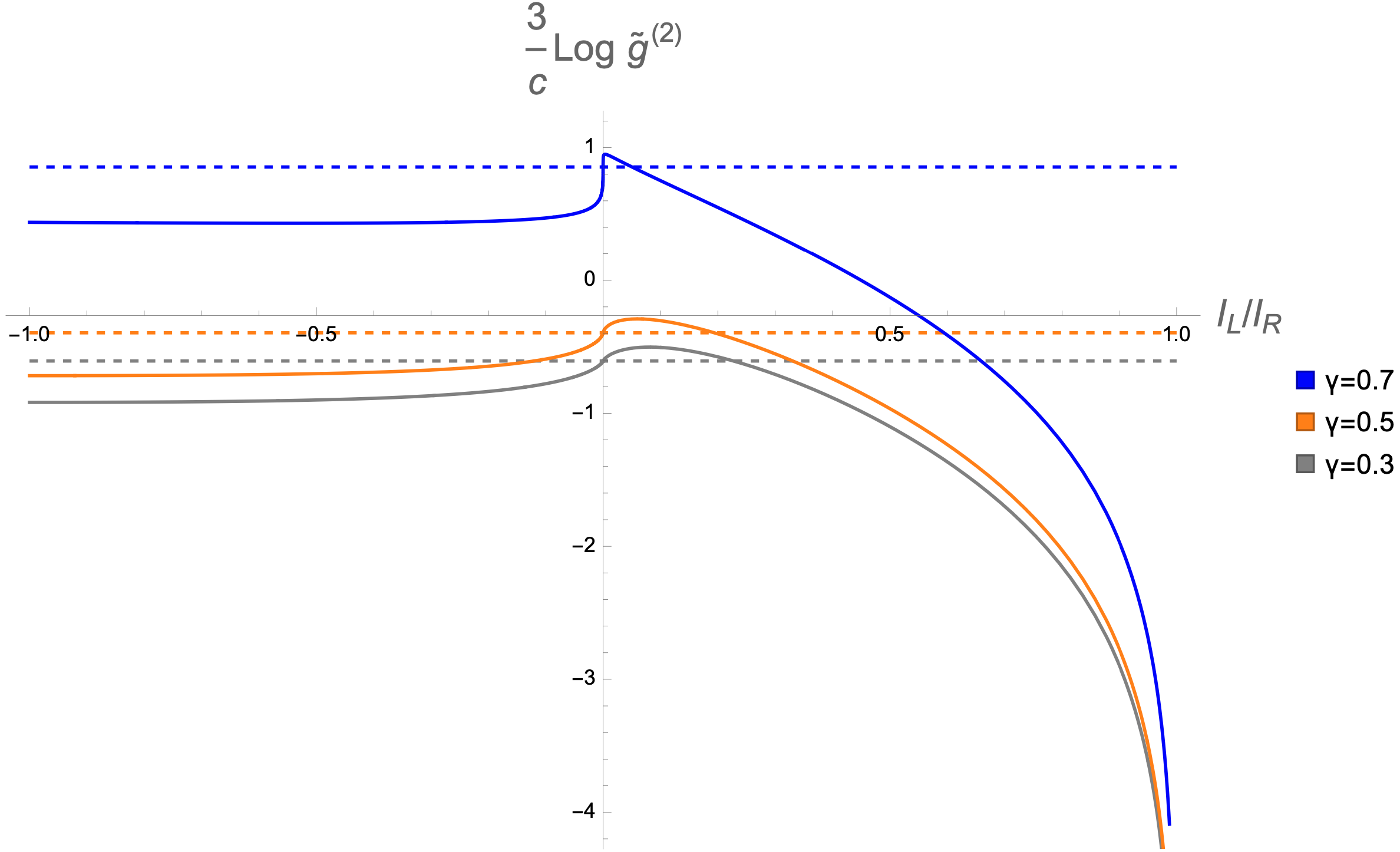}
    \caption{Plot of the scheme-independent contribution to the interface entropy \eqref{eq:logg2tilde} as a function of the ratio $ l_L/l_R$ for the Janus geometry, for different values of the parameter $\gamma$. Negative values correspond to crossing intervals and positive values to non/crossing intervals. The dashed lines correspond to the values in \eqref{eq:logg2tilde0}.}
    \label{fig:loggvslogratiodifferentγ}
\end{figure}

\begin{figure}[H] 
    \centering
   \includegraphics[width=0.7\textwidth]{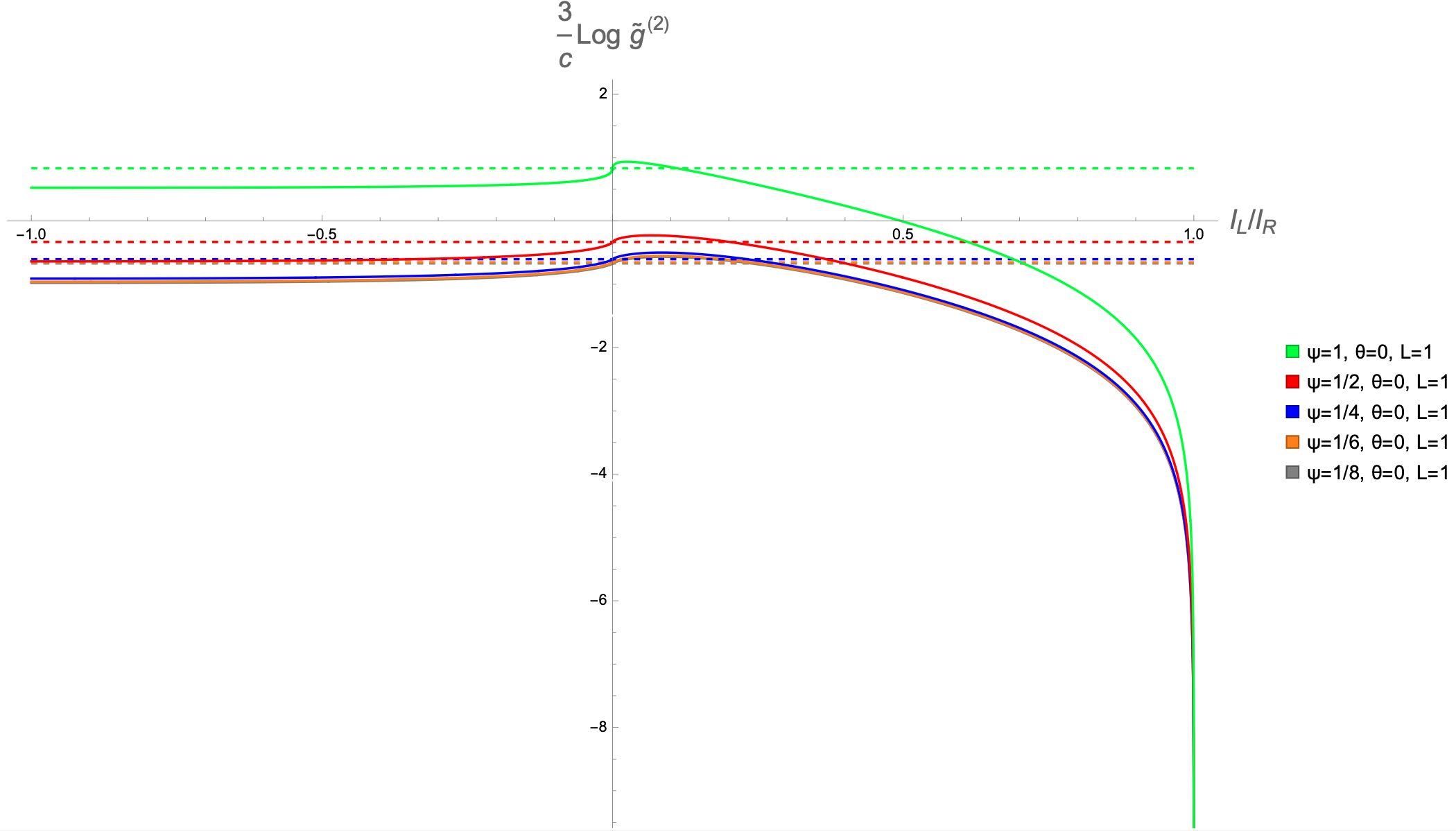}
    \caption{Plot of the scheme-independent contribution to the interface entropy \eqref{eq:logg2tilde} as a function of the ratio $ l_L/l_R$ for the super Janus geometry, for different values of the parameter $\psi$ and $\theta=0$, $L=1$. Negative values correspond to crossing intervals and positive values to non/crossing intervals. The dashed lines correspond to the values in \eqref{eq:logg2tilde0}.}
    \label{fig:loggvslogratiosuperJanus}
\end{figure}

The finite regularized part of the interface entropy and $\log(l_L/l_R)$ can be split in two contributions
\begin{equation}
\begin{split}
    &\log g_{\text{reg}}^{(2)}=\log g_{\text{IR}}^{(2)}+\log g_{\text{UV}}^{(2)},\\
     &\log \left( \frac{l_L}{l_R}\right)_{\text{reg}}=\log \left( \frac{l_L}{l_R}\right)_{\text{IR}}+\log \left( \frac{l_L}{l_R}\right)_{\text{UV}}.
    \end{split}
\end{equation}
The UV part takes the same form for crossing and non-crossing intervals
\begin{equation}
\begin{split}
    &\log g_{\text{UV}}^{(2)}=\frac{c}{3}\int_{r_0}^{\infty}dr\left(\frac{1}{\sqrt{1-c_s^2e^{-2A}R^{-2}}}-1 \right)+\log g,\\
    &\log \left( \frac{l_L}{l_R}\right)_{\text{UV}} =-2\int_{r_0}^\infty dr\,\frac{c_s e^{_-A}}{\sqrt{e^{2A}R^2-c_s^2}} \ .
\end{split}
\end{equation}
The IR contribution is
\begin{equation}
\begin{split}
    &\log g_{\text{IR}}^{(2)}=\frac{c}{3}\left[\int_{r_{\text{min}}}^{r_0}dr\left(\frac{1}{\sqrt{1-c_s^2e^{-2A}R^{-2}}}-1-\frac{e^{A_*}}{\sqrt{e^{2A_*}-c_s^2R^{-2}+2e^{2A_*}b_2(r-r_*)^2}} \right)-r_{\text{min}}\right]\ , \\
    &\log \left( \frac{l_L}{l_R}\right)_{\text{IR}} =-2\int_{r_{\text{min}}}^{r_0} dr\left(\frac{c_s e^{_-A}}{\sqrt{e^{2A}R^2-c_s^2}}-\frac{c_s R^{-1}e^{-A_*}}{\sqrt{e^{2A_*}-c_s^2R^{-2}+2e^{2A_*}b_2(r-r_*)^2}}\right)\ ,
\end{split}
\end{equation}
where $r_{\text{min}}=0$ for crossing intervals. Introducing these expressions in \eqref{eq:logg2complete}, we see that there are cancellations between the IR terms, in such a way that we can write the final expression in a way that is explicitly independent of $r_0$. Defining
\begin{equation}\label{eq:logg2tilde}
    \log \tilde g^{(2)}\equiv \frac{c \, e^{2A_*}R}{6c_s}\log \left( \frac{l_L}{l_R}\right)_{\text{reg}}+\log g_{\text{reg}}^{(2)},
\end{equation}
We obtain
\begin{equation}
    \log \tilde g^{(2)}=\frac{c}{3}\left[\int_{r_{\text{min}}}^\infty dr\left(\frac{1-e^{-2(A-A_*)}}{\sqrt{1-c_s^2e^{-2A}R^{-2}}}-1\right)-r_{\text{min}}\right]+\log g.
\end{equation}
In the $l_L/l_R\to 0$ limit it takes the simple form
\begin{equation}\label{eq:logg2tilde0}
    \log \tilde g^{(2)}=\frac{c}{3}\left[\int_{r_{\text{min}}}^\infty dr\left(\sqrt{1-e^{-2(A-A_*)}}-1\right)-r_{\text{min}}\right]+\log g.
\end{equation}
Where now for a crossing interval $r_{\text{min}}=0$, while for non-crossing intervals $r_{\text{min}}=r_*$. The difference of EEs is scheme-independent
\begin{equation}
    \log g_i\equiv \lim_{l_L/l_R\to 0} \left(S_A^{\text{crossing}}-S_A^{\text{non-crossing}}\right)=\frac{c}{3}\int_{0}^{r_*} dr\,\sqrt{1-e^{-2(A-A_*)}}.
\end{equation}
This could be interpreted as the contribution to the entropy from degrees of freedom at the interface. For Janus and super Janus geometries it vanishes, since $r_*=0$, while for the RS braneworld it takes a nonzero value 
\begin{equation}\label{eq:RSbraneJump}
    \log g_i=\frac{c}{3}\log\cosh r_*\ .
\end{equation}
We can evaluate numerically the finite scheme-independent contribution to the interface entropy $\log \tilde g^{(2)}$ \eqref{eq:logg2tilde} for the RS braneworld (Fig. \ref{fig:RSlogg2}), the Janus solution (Fig. \ref{fig:loggvslogratiodifferentγ}) and the super Janus solution (Fig. \ref{fig:loggvslogratiosuperJanus}).

\subsection{A $c$-theorem for $c_{\text{eff}}$} \label{sec::boundsbeta}

The value of $c_{\text{eff}}$ is directly related to the value of the pseudo-beta functions \eqref{eq:Bgc}, \eqref{eq:Bgnc} in the $l_L/l_R\to 0$ limit,
\begin{equation}
    \lim_{l_L/l_R\to 0} \BB= -\frac{c}{6}e^{A_*}=-\frac{c_{\text{eff}}}{6}\ .
\end{equation}
If we identify $C_{\text{eff}}\equiv -6\BB$ as an extension of $c_{\text{eff}}$ to intervals that do no have an endpoint at the interface, then it is possible to show that it obeys a ``$c$-theorem'' in the sense that it is a monotonically decreasing function of the ratio $l_L/l_R$. Notice that we do not perturb the ICFT in order to trigger a proper RG flow, but rather we constrain the functional dependence of $C_{\text{eff}}$ as a function of the ratio withing the ICFT. That $\BB$ is an increasing function with the ratio follows from the $g_{\text{eff}}$-theorem of \cite{Afxonidis:2024gne}. In order to see it, we will employ a setup on a timeslice with four crossing intervals as in \cite{Afxonidis:2024gne} (see figure 5 therein). Then, the SSA condition yields, in the conventions of \cite{Afxonidis:2024gne} (see their Eq. (3.7)),
\begin{equation}\label{eq::ineqF}
    \lim_{\rho\rightarrow 0}\left(\rho F'(\rho) \right)\leq \rho F'(\rho)\leq F'(1) \ ,\quad\quad F\left( \rho\right)\equiv\log g^{(2)}\left(\rho\right),\ \ \rho F'(\rho)=\BB \ .
\end{equation}
With $\rho=l_L/l_R$, and they showed that $F'(1)=0$. Thus $\log g^{(2)}$ is a monotonic function and the $g_{\text{eff}}$-theorem holds. Furthermore, we can rewrite \eqref{eq::ineqF} as
\begin{equation}\label{eq::boundsonbetag}
    -\frac{c_{\text{eff}}}{6}\leq \BB\leq 0 \ .
\end{equation}
This shows that $\BB\leq 0$, implying that $\log g^{(2)}$ monotonically decreases for all values of $\rho \in [0,1]$ and $\BB$ increases. 

Let us now show that $\BB$ is not only larger at $\rho=1$, but it is also monotonically increasing with $\rho$, which makes $C_{\text{eff}}$ monotonically decreasing. We will restrict to crossing intervals since it is technically simpler, but a similar derivation should be valid for non-crossing intervals. The equivalent statement we will prove is that the derivative
$d\BB / d \log g^{(2)}$ is negative. First, we differentiate \eqref{eq::firstderivative} one more time with respect to $c_s$
\begin{equation}\label{eq::secondderivatives}
      \frac{d^2\log \rho}{dc_s^2}=-\int_{-\infty}^{\infty}\frac{3c_se^A R^2}{\left( e^{2A}R^2-c_s^2\right)^{5/2}}<0 \ , \quad\quad \frac{d^2\log g^{(2)}}{dc_s^2}=\frac{c}{6}\int_{-\infty}^{\infty}\frac{\left( 2c_s^2+e^{2A}R^2\right)e^A R}{\left( e^{2A}R^2-c_s^2\right)^{5/2}}>0 \ .
\end{equation}
Then, the derivative of the pseudo-beta function is
\begin{align}
    \frac{d\BB}{d\log g^{(2)}}&=\frac{dc_s}{d\log g^{(2)}}\frac{1}{\left( \frac{d\log \rho}{dc_s}\right)^2}\left(\frac{d\log \rho}{dc_s}\frac{d^2\log g^{(2)}}{dc_s^2}-\frac{d\log g^{(2)}}{dc_s}\frac{d^2\log \rho}{dc_s^2} \right) \nonumber \\
    &=-\frac{c\,R}{6}\frac{dc_s}{d\log g^{(2)}}\frac{1}{\left( \frac{d\log \rho}{dc_s}\right)^2}\left(\int_{-\infty}^{\infty}\frac{e^AR}{\left( e^{2A}R^2-c_s^2\right)^{3/2}} \right)^2<0 \ ,
\end{align}
where we have employed (\ref{eq::firstderivative}) and (\ref{eq::secondderivatives}).

\begin{figure}[H] 
    \centering
    \includegraphics[width=0.7\textwidth]{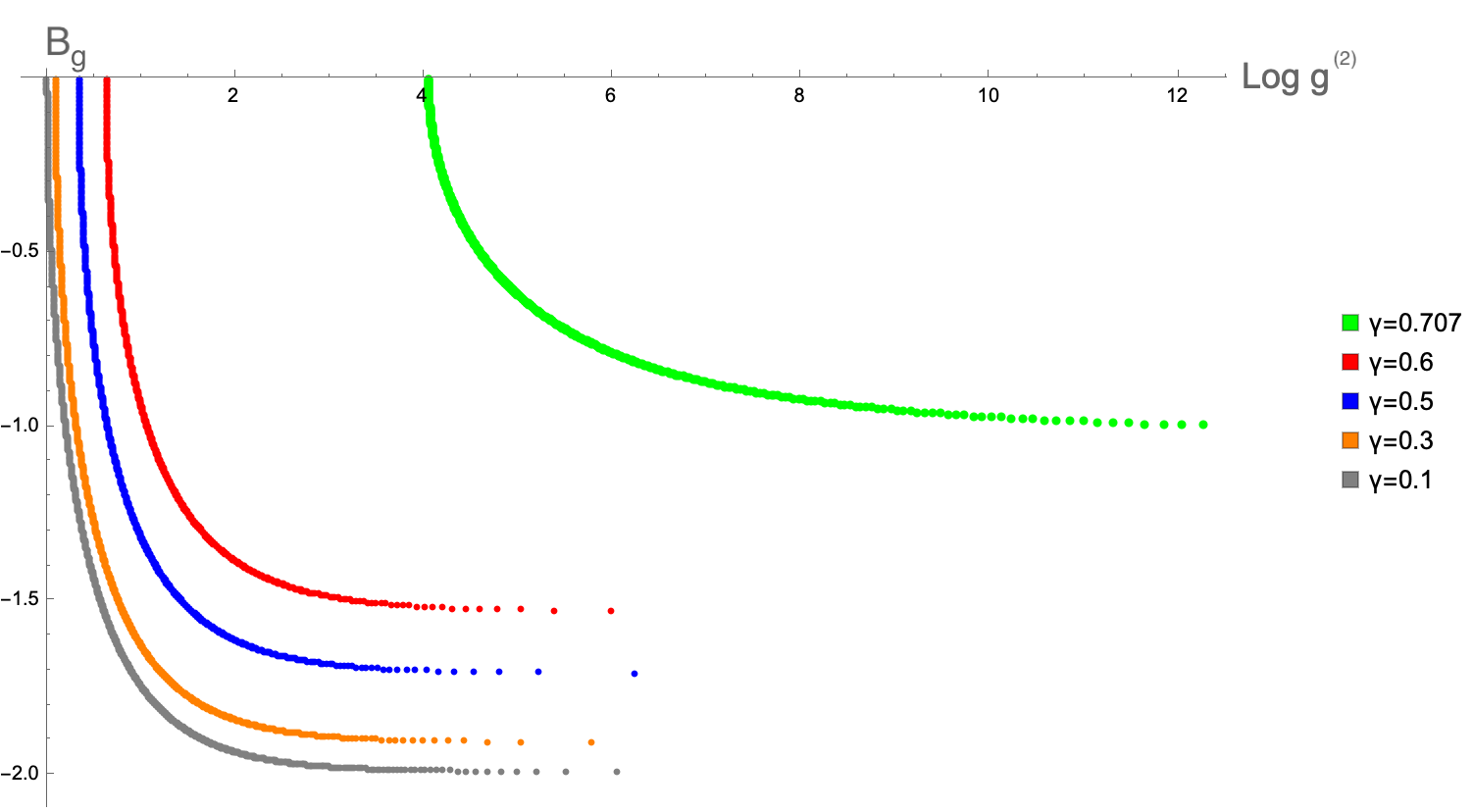}
    \caption{Plot of the $\BB$ function with respect to $\log g^{(2)}$ for the Janus geometry for different values of the $\gamma$.}
    \label{fig:betaloggJanus}
\end{figure}

\begin{figure}[H] 
    \centering
    \includegraphics[width=0.7\textwidth]{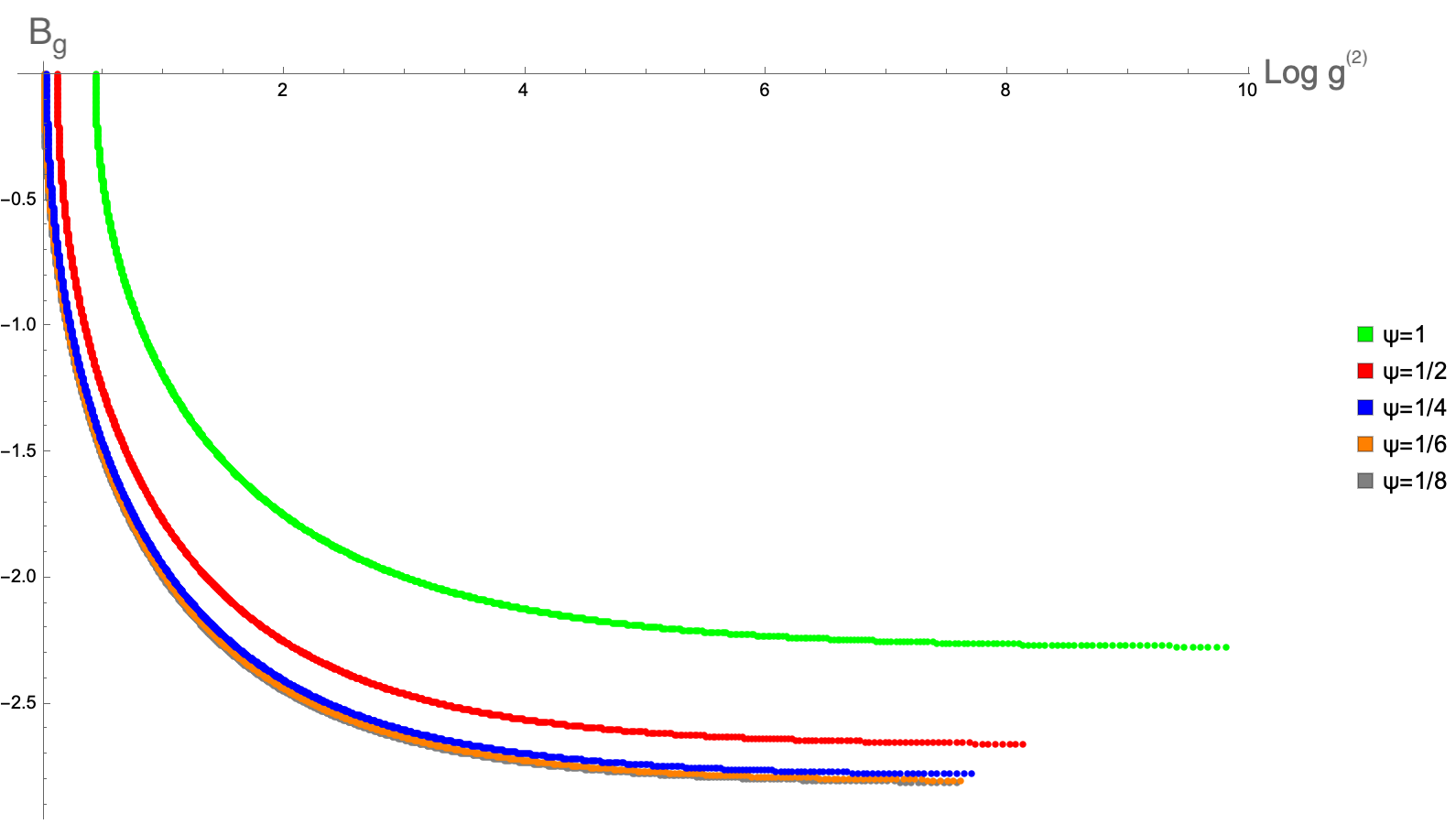}
    \caption{Plot of the $\BB$ function with respect to $\log g^{(2)}$ for the super Janus geometry for different values of the $\psi$ and $\theta=0$ and $L=1$.}
    \label{fig:betaloggsuperJanus}
\end{figure}

We have checked numerically these results for the Janus and super Janus geometries. The behavior of the pseudo-beta function $\BB$ for the standard Janus and super Janus for different values of the parameters $\gamma$ and $\psi$ can be seen in figures \ref{fig:betaloggJanus} and \ref{fig:betaloggsuperJanus} respectively.

From figures \ref{fig:betaloggJanus} and \ref{fig:betaloggsuperJanus} we see that $\BB$ is monotonically decreasing with $\log g^{(2)}$.  Note that when $\BB=0$, the interface entropy acquires the value of the boundary entropy number $\log g$ for symmetric intervals around the interface. On the other hand, for large values of $\log g^{(2)}$, $\BB$ approaches the value $-c_{\text{eff}}/6$.

\begin{figure}[H] 
    \centering
    \includegraphics[width=0.8\textwidth]{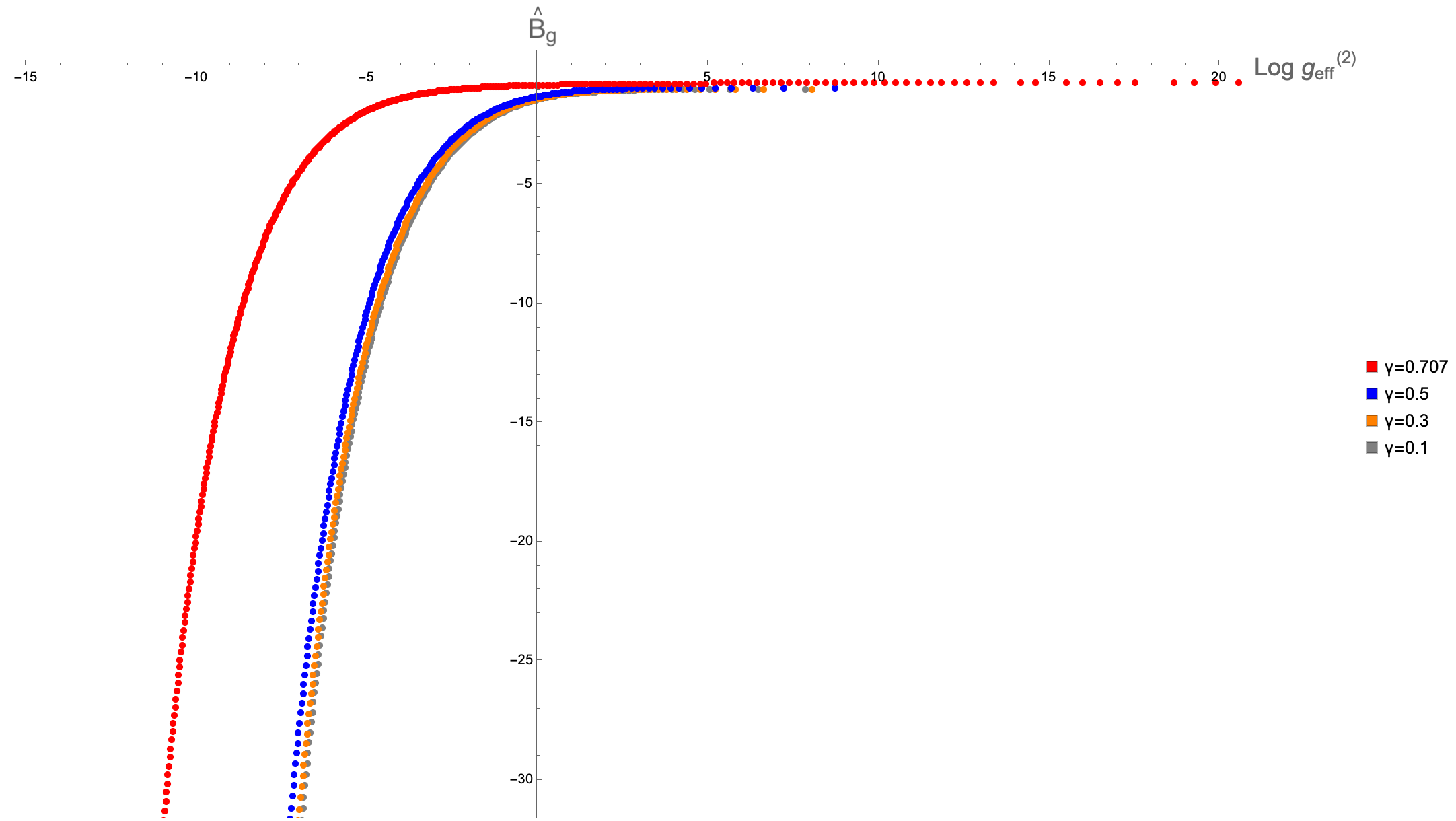}
    \caption{${\BB}$ for the Janus geometry with respect to the interface entropy $\log g^{(2)}$ for various values of $\gamma$.}
    \label{fig:betavsloggjanusnoninterface}
\end{figure}

\begin{figure}[H] 
    \centering
    \includegraphics[width=0.8\textwidth]{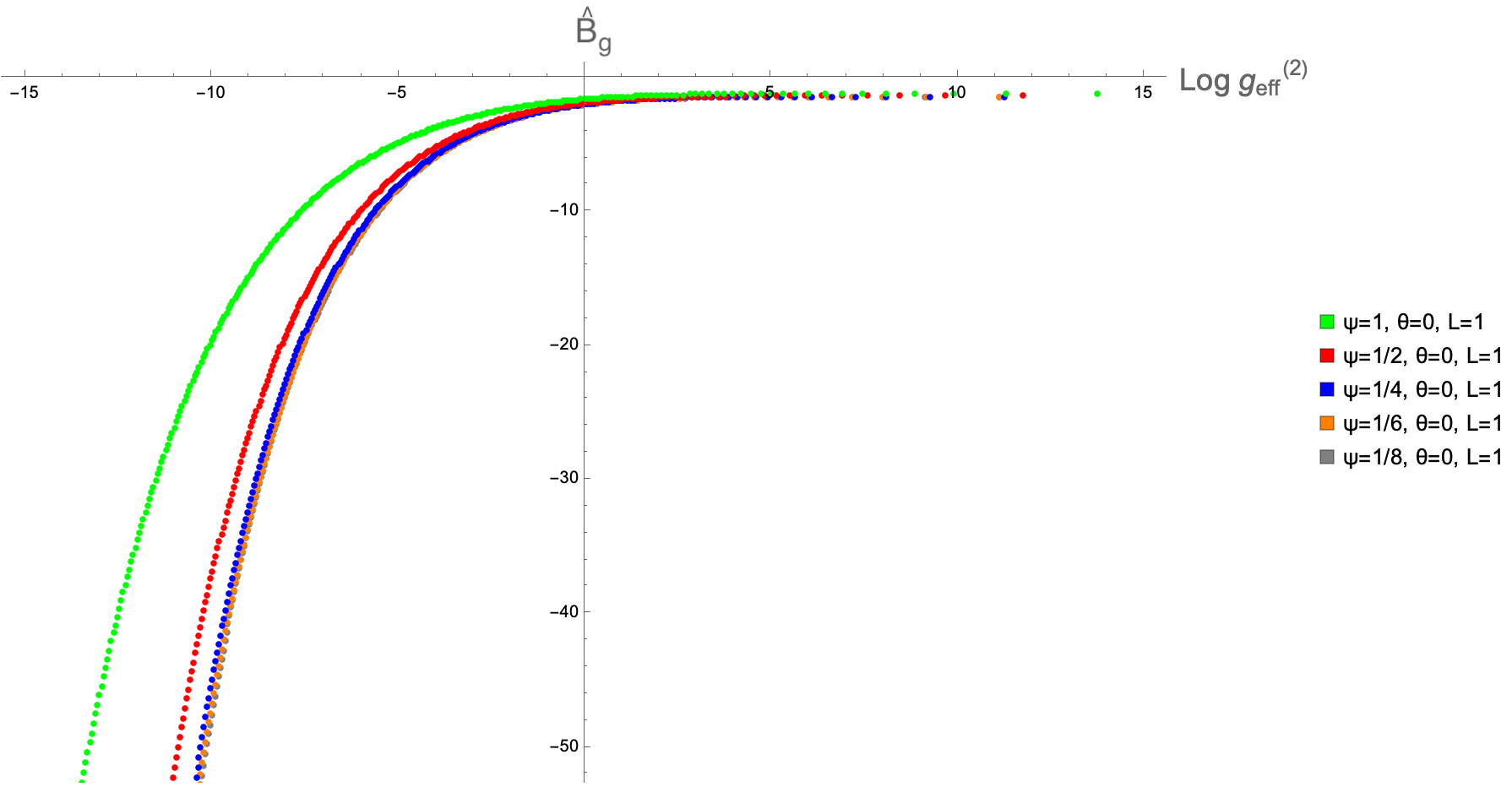}
    \caption{${\BB}$ for the super Janus geometry with respect to the interface entropy $\log g^{(2)}$ for $\theta=0$, $L=1$ and various values of the dilaton $\psi$.}
    \label{betavsloggsuperjanusnoninterface}
\end{figure}

We also compute the $\BB$ for non-crossing intervals numerically, for Janus and super Janus in figure \ref{fig:betavsloggjanusnoninterface} and figure \ref{betavsloggsuperjanusnoninterface} respectively.
We confirm that the pseudo-beta function $\BB$ is always negative. $\BB$ approaches the value $-c_{\text{eff}}/6$ for large values of $\log g^{(2)}$, which is the region where the interface entropy has a logarithmic divergence as it hits the interface, $l_L/l_R \rightarrow 0$.

\section{Asymmetric ICFTs and BCFTs}\label{sec::unequalcentralcharges}

So far we have discussed only ICFTs with equal central charges on both sides of the interface $c_L=c_R=c$. One can also consider asymmetric ICFTs where $c_L \neq c_R$. In the extreme limit  where one of the central charges vanish, say $c_L\to 0$, the theory becomes effectively a BCFT living on the right side of the interface, which now has become a boundary.

\begin{figure}[H] 
    \centering
    \includegraphics[width=0.75\textwidth]{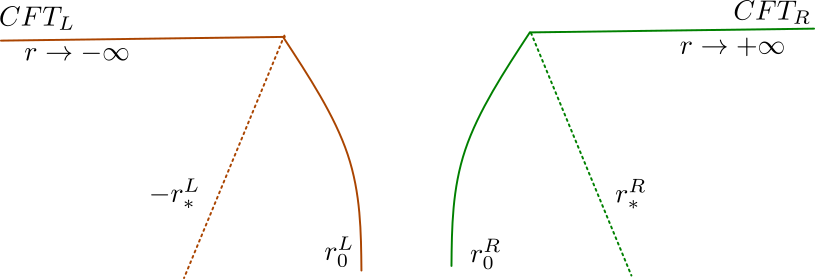}
    \caption{The bulk geometry of an asymmetric interface within the RS setup. The left and right $AdS_3$ spaces are cut along different slices $r_0^L$ and $r_0^R$ and glued together through the Israel junction conditions. For $r_0^L=r_0^R=0$, the minima on each side is achieved at $r=r_*^R,-r_*^L$.} 
    \label{fig:TwoAdSPatches}
\end{figure}

The entanglement entropy in an asymmetric ICFT will take the form
\begin{subequations}   
\begin{align}
\label{eq:SAcasym}    S_A^c & =\frac{c_L}{6}\log \frac{2l_{L}}{\epsilon}+\frac{c_R}{6}\log \frac{2l_{R}}{\epsilon}+\log g^{(2)}_c \ ,\\
     S^{nc}_A&=\frac{c_R}{6}\log \frac{2l_L}{\epsilon}+\frac{c_R}{6}\log \frac{2l_R}{\epsilon}+\log g^{(2)}_{nc}\ .
    \end{align}
\end{subequations}
Where $S_A^c$ is the EE for a crossing interval and $S_A^{nc}$ for a non-crossing interval to the right of the interface.  In the holographic dual description we can think of this as having two different AdS radii $R_L$ and $R_R$ for the geometry to the left and to the right of the interface. This type of geometry can be constructed using the RS braneworld setup, as in figure \ref{fig:TwoAdSPatches}.

The geometry is such that, on each side of the interface, we introduce a dimensionful radial coordinate by rescaling the dimensionless radial coordinate appearing in \eqref{eq::defectmetric} as
\begin{equation}
    ds^2=dr^2+R_i^2 e^{2A_i(r)}ds^2_{AdS_2} \ ,
\end{equation}
where $i=L,R$ labels the left and the right wedges. We keep the $r<r_0^L$ region of an $AdS_3$ space of radius $R_L$ and the $r>r_0^R$ region of another $AdS_3$ space of radius $R_R$.  We glue these two spaces at their boundaries by introducing a brane of some definite tension and applying the Israel junction conditions \cite{Israel:1966rt}. The first condition forces the induced metric on the brane to be the same for the left and the right parts of the $AdS_3$

\begin{equation} \label{eq::Israel1}
    R_L (e^{A_0})_{L}= R_R (e^{A_0})_{R}\,.
\end{equation}
A possible choice is to take 
\begin{equation}
    e^{A_L(r)}=\cosh\left(\frac{r-r_0^L+r_*^L}{R_L}\right), \quad e^{A_R(r)}=\cosh\left(\frac{r-r_0^R- r_*^R}{R_R}\right),
\end{equation}
such that
\begin{equation}
    \frac{\cosh\left(\frac{r_*^R}{R_R}\right)}{\cosh\left(\frac{r_*^L}{R_L}\right)}=\frac{R_L}{R_R}\,.
\end{equation}
The second Israel junction condition is related to the tension of the brane $T$, and enforces it to be positive. For the sake of comparison we will locate the brane at $r_0^L=r_0^R=0$. With this choice, the boundary entropy \eqref{eq:gisym} becomes
\begin{equation}
    \log g=\frac{r_*^L+r_*^R}{4G}\,.
\end{equation}

We recover reflection symmetry in the symmetric ICFT, $r_*^L=r_*^R=r_*$ when $R_L=R_R=R$. The BCFT limit $c_L\to 0$ ($R_L\to 0$) can also be found formally by taking  $r_*^L/ R_L\to +\infty$ while keeping $R_L \cosh (r_*^L/R_L)$ fixed. In this case the spacetime is extended on $r\geq 0$, with the boundary at $r=0$. This is simply related to the original holographic BCFT proposal \cite{Takayanagi:2011zk} by a shift in the $r$ coordinate.

In the following we will re-analyze some of the results we obtained for symmetric ICFTs in this scenario.

\subsection{Crossing intervals in asymmetric ICFTs}

For crossing intervals the EE is usually presented as
\begin{equation}
      S_A^c  =\frac{c_L+c_R}{6}\log \frac{l_{L}+l_R}{\epsilon}+\log g_c^{(1)}\ .
\end{equation}
The relation with the interface entropy in \eqref{eq:SAcasym} is
\begin{equation}
\label{eq:geff1geff2conv}
    \log g_c^{(1)}= \log g_c^{(2)} + \frac{c_L}{6} \log \frac{2 l_L}{l_L+l_R} + \frac{c_R}{6} \log \frac{2l_R}{l_R + l_L}\ .
\end{equation}
In this case the minimal length geodesics that determine the EE of the interval explore the two regions with different $AdS$ radii, so we need to introduce a generalization of the formulas derived previously, taking into account that the radius and warp factor are no longer reflection symmetric across the interface. 

Considering the equation for the minimal length geodesic \eqref{eq::diffeq}, and demanding that the geodesic solution crosses smoothly through the brane, gives the condition 
\begin{equation}\label{eq:matchingcond}
    \frac{x_L'}{x_L}\Bigg|_{r=0}= \frac{x_R'}{x_R}\Bigg|_{r=0}\quad \Rightarrow\ \quad c_s^L=c_s^R= c_s\ .
\end{equation}

The formulas for the $l_L/l_R$ ratio \eqref{eq::logratioequation} and the interface entropy \eqref{eq::loggicc} generalize to
\begin{subequations}\label{eq:logsasym}
    \begin{align}
\log\left( \frac{l_L}{l_R}\right)=&-\int_{-\infty}^0 \frac{dr}{R_L} \frac{c_s e^{-A_L}}{\sqrt{e^{2A_L}R_L^2-c_s^2}} -\int_{0}^\infty \frac{dr}{R_R} \frac{c_s e^{-A_R}}{\sqrt{e^{2A_R}R_R^2-c_s^2}},\\
 \notag \log g_c^{(2)} =& \log g+\frac{c_L}{6}\int_{-\infty}^{0} \frac{dr}{R_L}\left(\frac{1}{\sqrt{1-c_s^2e^{-2A_L}R_L^{-2}}}-1 \right)\\ & +\frac{c_R}{6}\int_0^\infty \frac{dr}{R_R}\left(\frac{1}{\sqrt{1-c_s^2e^{-2A_R}R_R^{-2}}}-1 \right)
 \ .
    \end{align}
\end{subequations}

Let us now analyze the limits $l_L/l_R\rightarrow 0$ and
$l_R/l_L\rightarrow 0$. Since the conserved quantity is the same on
both sides of the interface, $c_s^L=c_s^R\equiv c_s$, the reality of
the geodesic solution requires
\begin{equation}
    |c_s| < \min(R_L,R_R) \, ,
\end{equation}
where we have used the fact that the minimum value of the warp factor
on both sides is $e^{A_{L,R}}=1$.
If $R_L<R_R$, the limiting value is therefore
\begin{equation}
    |c_s|\rightarrow R_L \, .
    \label{eq:cs_limit_left}
\end{equation}
In this limit, the denominator of the left contributions in
\eqref{eq:logsasym} vanishes at the minimum of
the left warp factor, $r=-r_\ast^L$, while the contribution from the
right side remains finite, since
\begin{equation}
    \frac{|c_s|}{R_R}
    \rightarrow
    \frac{R_L}{R_R}<1 \, .
\end{equation}
Therefore, only the integrals over the left side of the interface
develop a logarithmic divergence. Whether this limit corresponds to
$l_L/l_R\rightarrow 0$ or $l_R/l_L\rightarrow 0$ depends on the sign
chosen for $c_s$, corresponding to the two possible branches of the
geodesic solution.

For concreteness, let us assume $R_L<R_R$, and study the limit $c_s\to R_L$. Taking $l_L\approx \lambda \epsilon$, $l_R\approx l$, the EE for a crossing interval becomes
\begin{equation}
    S_A^c\approx \frac{c_R}{6}\log \frac{l}{\epsilon}+\frac{c_L}{6}\log(2\lambda)+\frac{c_R}{6}\log(2)+\log g^{(2)}_c\, .
\end{equation}
As in the symmetric case, there are logarithmically divergent contributions in the integrals appearing \eqref{eq:logsasym}, close to the points where the warp factors takes its minimal value $e^{A_L(-r_L^*)}=1$. 
\begin{equation}
    \log g^{(2)}_{\text{div}}=\frac{c_L}{6}{\cal I}_L, \quad \log\left(\frac{l_L}{l_R} \right)_{\text{div}}=-\frac{c_s}{R_L} {\cal I}_L \ ,
\end{equation}
where, picking a suitable value for $\lambda$ and $r_0> r_*^L,r_*^R$,
\begin{equation}\label{eq::logglogratioregdiv2asym}
    \mathcal{I}_L=\int_{-r_0}^{0}\frac{dr}{R_L\sqrt{1-c_s^2R_L^{-2}+c_s^2R_L^{-4}(r+r_*^L)^2}} \ .
\end{equation}
Using that, in the limit $c_s\to R_L$, we arrive at
\begin{equation}
   \log g^{(2)}_{\text{div}} \sim -\frac{c_L}{6}  \log\left(\frac{l_L}{l_R} \right)_{\text{div}}.
\end{equation}
Thus, for a crossing interval,
\begin{equation}
    c_{\text{eff}}^{R_L<R_R}=c_L \ .
\end{equation}
An analogous analysis for $R_R<R_L$ shows that when we take the limit $c_s\to R_R$, we obtain
\begin{equation}
    c_{\text{eff}}^{R_R<R_L}=c_R \ .
\end{equation}
Therefore, for crossing intervals,
\begin{equation}\label{eq:ceffasym}
    c_{\text{eff}}=\text{min}(c_L,c_R)\ .
\end{equation}
This is consistent with the bound found in \cite{karch2023universalityeffectivecentralcharge,karch2024universalboundeffectivecentral}.

\subsection{Non-crossing intervals in asymmetric ICFTs}

When the AdS radii on each side of the brane are different, there can be several possible competing configurations for non-crossing intervals \cite{Anous_2022}. One type of configuration is like the one we have already discussed, where the minimal length geodesic remains on the same side of the interface. Other configurations are such that the geodesic goes through the brane to the region with different AdS radius and then returns.  

For the first type, all formulas for non-crossing intervals remain unchanged, since they only explore a region with a fixed $AdS$ radius. In particular, they will have the same monotonicity properties that we derived previously and when one endpoint of the interval reaches the interface, there will be a logarithmically divergent contribution from the interface entropy with a coefficient proportional to $c_{\text{eff}}^{L,R}\propto R_{L,R} e^{A_{L,R}(r_*^{L,R})}=R_{L,R}$, depending on which endpoint is taken to the interface. 

The other type of geodesic is composed of three pieces. Two of them remain on the same side of the brane and extend from the endpoints of the interval at the boundary to the brane.  The third piece lies at the other side of the brane and joins the other two forming a continuous and smooth curve. Demanding that the geodesic crosses smoothly the brane leads to the same matching conditions as for crossing intervals \eqref{eq:matchingcond}, so all three pieces are determined by the same value of $c_s$. In order for this configuration to be possible, the value of $c_s$ must be smaller than the AdS radius on the side where the endpoints of the interval are located, so they can reach the brane before turning around. On the other hand, $c_s$ must be larger than the AdS radius at the other side, so the third piece has a turnaround point and can return to the brane. This implies two things: First, these configurations only exist when the interval is on the side of the interface with larger central charge (larger AdS radius). Second, when they exist, the logarithmically divergent contribution is determined by the AdS radius at the location of the turning point, which is the smaller radius.

The situation is then the following. When the interval is on the side of smaller central charge, only the usual non-crossing configuration exist, and $c_{\text{eff}}$ is equal to the smaller central charge. On the other hand, when the interval is on the side of larger central charge, we have two competing configurations. When one of the endpoints of the interval approaches the interface, both configurations have a logarithmic divergence, but the coefficient is smaller for the configuration where the geodesic goes through the brane. Thus, this configuration has smaller length and the entanglement entropy will be determined by it. It then follows that also for non-crossing intervals  \eqref{eq:ceffasym} is also satisfied.

\section{Discussion}\label{sec::discussion}

We have studied in detail the boundary entropy function associated to general intervals in holographic $1+1$-dimensional ICFTs, including the less explored case of intervals that do not cross the interface. A difference from CFTs that do not have interfaces is that the entanglement entropy acquires a finite ``boundary entropy'' contribution that exhibits interesting features depending on the geometry of the interval. Another seemingly unrelated aspect of the entanglement entropy in a ICFT is that the coefficient of the logarithmic divergence for an interval with an endpoint at the interface changes to an effective central charge  $c_{\rm eff}$ smaller than the central charge of the bulk CFT. One of our main results is to show that these two features are not independent.

The main quantity dictating the properties of the boundary entropy function is the ratio $l_L/l_R$ of the distance of the endpoints of the interval to the interface.  We have shown the explicit relation between the boundary entropy function and the effective central charge by taking the limit where one endpoint of the entangled region hits the interface. In this limit a logarithmic divergence appears in the boundary entropy function, that precisely accounts for the effective central charge. 

We have also argued that strong subadditivity of entanglement entropy requires having a non-zero boundary entropy function even for intervals that do not cross the interface, in such a way that the same value of the effective central charge is found independently of whether the limit where an endpoint hits the interface is taken for crossing or non-crossing intervals. Consequently, a finite scheme-independent interface entropy can be defined by subtracting the EE of crossing and non-crossing intervals in this limit, which we found is vanishing for Janus and super Janus geometries but non-zero for the RS braneworld construction. In the opposite limit $l_L/l_R\to 1$, where the crossing interval becomes symmetric around the interface, the interface entropy becomes the boundary entropy. In this limit, the non-crossing interval collapses to a point, and the interface entropy vanishes.

Additionally, we have defined a ``c-function'' for an analog of an RG flow depending on the ratio $l_L/l_R$. We remark that this is not a proper RG flow, since the ICFT is unchanged and only the shape of the entangled region is modified. Nevertheless, there is a quantity $C_{\text{eff}}$ that is monotonically decreasing from $C_{\text{eff}}=c_{\text{eff}}$ when one of the endpoints is at the interface to $C_{\text{eff}}=0$ when the interval is symmetric around the interface.

There are several avenues that would be interesting to explore in light of our results. In the first place, our analysis has been restricted to holographic duals to ICFTs, it would be interesting to have a purely field theory derivation of the boundary entropy limits giving the effective central charge. For non-crossing intervals in particular, there is no analysis of the boundary entropy in ICFTs to the best of our knowledge. A study of the Strong Subadditivity conditions may lead to similar conclusions than those derived from the holographic models.

Another interesting question is whether one can find a $c_{\text{eff}}$-theorem that characterizes RG flows at the interface. This would involve understanding how does the effective central charge evolve with respect to a relevant deformation of the ICFT. On a similar note, one can study the evolution of the boundary entropy function with respect to an interface RG flow. For a symmetric interval around the interface, this would in principle reproduce the entropic proof of of the $g$-theorem \cite{harper2024gtheoremstrongsubadditivity}. Note that proofs of a $c$-theorem based on entanglement entropy make use of strong subadditivity for boosted intervals \cite{Casini:2004bw,harper2024gtheoremstrongsubadditivity}. However, in the presence of an interface there is a preferred frame, so the relation between the entropies of an interval and its boosted version may be more complicated than in an ordinary globally Lorentz-invariant CFT. A proper calculation of the entanglement entropy in holographic ICFTs thus requires using an extension of the Ryu-Takayanagi prescription to boosted intervals like the one proposed in \cite{Hubeny:2007xt}. Alternatively, one can study the temperature dependence of the boundary entropy \cite{Friedan:2003yc}, that has also been studied in holographic models of BCFTs and ICFTs e.g. \cite{Takayanagi:2011zk,Erdmenger:2015spo,Erdmenger:2015xpq,Liu:2024oxg}.

Another very interesting question is how much of these results carries over to higher dimensions. As already noted in \cite{karch2023universalityeffectivecentralcharge,karch2024universalboundeffectivecentral} the holographic construction carries over to higher dimensions without any major modifications, but the entanglement entropy now  has more singular terms and develops dependence on the shape of the entangling region, so it is less obvious what aspects to focus on.

 \acknowledgments
We would like to thank Elias Kiritsis, Chitraang Murdia and Chris Rosen for useful discussions. We thank Mianqi Wang for pointing out an error in an earlier version. The work of E.A. is supported by the Severo Ochoa fellowship PA-23-BP22-170 and the work of I.C.B. is supported by the Severo Ochoa fellowship  NAC-AT-PUB-ASV-2025 BP24-116. The work of E.A., I.C.B. and C.H. are partially supported by the Spanish Agencia Estatal de Investigaci\'on and Ministerio de Ciencia, Innovaci\-on y Universidades through the grant PID2021-123021NB-I00. The work of AK is supported in part by the U.S. Department of Energy under Grant No. DE-SC0022021 and a grant from the Simons Foundation (Grant 651678, AK).  E.A acknowledges the INGENIUM Alliance of European Universities for providing the opportunity to spend training time at the University of Crete and  would like to thank the University of Crete for the kind hospitality.

\bibliographystyle{JHEP}
\bibliography{biblio.bib}
\end{document}